\documentclass[journal]{IEEEtran}
\ifCLASSINFOpdf \else \fi

\usepackage{multirow}
\usepackage{bbold}
\usepackage{ifpdf}
\usepackage{ulem} 
\usepackage{subcaption}%
\usepackage[cmex10]{amsmath}
\usepackage{amssymb}
\usepackage{verbatim}
\usepackage{algorithm}
\usepackage{algorithmic}
\usepackage{array}
\usepackage{latexsym}
\usepackage{color}
\usepackage{textgreek}
\usepackage{subscript}
\usepackage{rotating}
\usepackage{booktabs,multirow}
\usepackage{mdframed}	
\usepackage{tabularx}
\usepackage{amsmath}
\usepackage{makecell}
\usepackage{tabto}
\usepackage{mathrsfs}
\usepackage{url}
\usepackage{cite}
\usepackage{listings}
\usepackage{array}
\usepackage[table]{xcolor}
\usepackage{ragged2e}
\usepackage{booktabs}

\usepackage{graphicx}

\usepackage[cmex10]{amsmath}
\usepackage{amssymb}
\usepackage{bbold}
\usepackage{float}
\usepackage[cmex10]{amsmath}
\usepackage{amssymb}
\usepackage{verbatim}
\usepackage{array}
\usepackage{latexsym}
\usepackage{color}
\usepackage{tabularx} 
\usepackage{textgreek}
\usepackage{subscript}
\usepackage{rotating}
\usepackage{booktabs,multirow}
\usepackage{mdframed}	
\usepackage{tabularx}
\usepackage{amsmath}
\usepackage{makecell}
\usepackage{tabto}

\usepackage{mathrsfs}
\usepackage{url}
\usepackage{cite}
\usepackage{listings}
\usepackage{array}
\usepackage[table]{xcolor}

\newcommand{\noteblue}[1]{\textcolor{black}{{\bf #1}}}
\definecolor{pinkpurple}{rgb}{0.6, 0.1, 0.9} 
\usepackage{algorithmic}
\usepackage{array}
\usepackage{algorithm}
\usepackage{soul}
\sethlcolor{green}
\usepackage{url}
\usepackage{pifont}
\usepackage{xcolor}
\newcommand{\cmark}{\textcolor{green}{\ding{51}}} 
\newcommand{\xmark}{\textcolor{red}{\ding{55}}}   
\usepackage{multirow}
\usepackage{graphicx} 

\begin{document}

\title{Joint Optimization of Completion Ratio and Latency of Offloaded Tasks with Multiple Priority Levels in 5G Edge}

\author{
	~Parisa Fard Moshiri, Murat~Simsek,~Burak Kantarci
	\thanks{
	The authors are with the School of Electrical Engineering and 
        Computer Science at the University of Ottawa, Ottawa, ON, K1N 6N5, Canada.
		E-mail: \{ parisa.fard.moshiri,murat.simsek,burak.kantarci\}@uottawa.ca }

}

\markboth{IEEE Transactions on Network and Service Management}
{Submitted \MakeLowercase{\textit{.}}: }

\maketitle
 \thispagestyle{empty}
  \pagestyle{empty}
\begin{abstract}
 Multi-Access Edge Computing (MEC) is widely recognized as an essential enabler for applications that necessitate minimal latency. However, the dropped task ratio metric has not been studied thoroughly in literature. Neglecting this metric can potentially reduce the system's capability to effectively manage tasks, leading to an increase in the number of eliminated or unprocessed tasks. This paper presents a 5G-MEC task offloading scenario with a focus on minimizing the  dropped task ratio, computational latency, and communication latency. We employ Mixed Integer Linear Programming (MILP), Particle Swarm Optimization (PSO), and Genetic Algorithm (GA) to optimize the latency and dropped task ratio. We conduct an analysis on how the quantity of tasks and User Equipment (UE) impacts the ratio of dropped tasks and the latency. 
The tasks that are generated by UEs are classified into two categories: urgent tasks and non-urgent tasks. The UEs with urgent tasks are prioritized in processing to ensure a zero-dropped task ratio. Our proposed method improves the performance of the baseline methods, First Come First Serve (FCFS) and Shortest Task First (STF), in the context of 5G-MEC task offloading. Under the MILP-based approach, the latency is reduced by approximately 55\% compared to GA and 35\% compared to PSO. The dropped task ratio under the MILP-based approach is reduced by approximately 70\% compared to GA and by 40\% compared to PSO.
\end{abstract}

\IEEEpeerreviewmaketitle

\section{Introduction}
The increasing demands and the bottleneck in the network core to access cloud services have called for Mobile Edge Computing (MEC) in the current technological environment \cite{nen23}. With the rapid expansion in the number of connected devices and the volume of data they generate, conventional cloud computing models, which include processing data in data centres, are facing challenges due to bottlenecks in the backbone network, higher latency, and bandwidth limitations\cite{wu22}. MEC tackles these difficulties by deploying computing resources in close proximity to the data generation site, namely at the network edge\cite{feng22}. The close proximity between components significantly decreases the latency, allowing for quicker data processing and decision-making, which is essential for applications such as self-driving cars, augmented reality, Internet of Things (IoT) devices, and intelligent urban infrastructure \cite{pong23}. Some of the edge functionalities are used for real-time communication protocol assistance, such as self-driving cars \cite{wu22}.

The integration of MEC in a 5G network offers numerous revolutionary advantages \cite{rana21}. Firstly, the expanded bandwidth capabilities of 5G enable more efficient processing of larger volumes of data \cite{elgendy20}. This is especially crucial in situations where numerous IoT devices are networked, necessitating the simultaneous transfer of significant volumes of data\cite{mustafa22}. Furthermore, the crucial aspect of 5G's ultra-low latency is significant in minimizing response times in critical applications, including autonomous driving and smart grid control \cite{pham20}. By processing data at the edge, in close proximity to its source, and utilizing the low latency of 5G, these applications can operate with greater reliability and efficiency \cite{islam21}.

To tackle the difficulties posed by the restricted processing capabilities and limited resources of smart devices, the industry has introduced the concept of computational offloading as part of the MEC framework \cite{singh23}. This method entails assigning computationally intensive tasks to dedicated MEC servers that possess adequate computational capabilities \cite{ali23}. Once these tasks are executed on the MEC server, the results are then accessed by the UE. Additionally, developing applications in 5G heavily depends on computational offloading technologies to efficiently provide services to UEs\cite{hou22}.

When UE tasks are offloaded to the MEC server, there are several factors that affect communication and computational latency together with  the task completion rate. Tasks are placed in a memory queue during the offloading process, which may result in execution latency. Therefore, tasks that exceed their predetermined deadline and remain in the queue are considered dropped. The effectiveness of the offloading process is highly dependent on the use of offloading methods and the allocation of resources\cite{yang23}. This encompasses the optimization of resource utilization\cite{farooq21}, reduction of latency \cite{fard24}, energy conservation \cite{li21}, enhancement of network utilization\cite{hsu22}, and achieving cost-efficiency \cite{islam21}. Additionally, in scenarios where time-sensitive tasks are involved, there may be UEs whose tasks need urgent or high-priority processing, which significantly influences the task offloading strategies and resource management in MEC \cite{farooq21}.

Based on a comprehensive review of published works on this topic, it appears that only a few studies have included the concept of dropped task ratio as a metric for optimization, in addition to other metrics such as latency \cite{feng22}\cite{pong23}\cite{elgendy20}. By including this metric in the optimization process, it is possible to come up with more effective decisions in terms of overall system efficiency and UE satisfaction \cite{fard24}. This guarantees that tasks are not only executed efficiently (with low latency) but also have a smaller chance of being dropped due to inadequate or missed deadlines. Therefore, when trying to achieve the minimum latency, it is crucial to consider the dropped task ratio.

Previously in \cite{fard24}, we minimized computational delay and dropped task ratio using MILP and GA without considering the communication latency or priority levels for the tasks. In this study, we utilize optimization techniques to minimize not only computational and communication latencies but also the often-overlooked metric of dropped task ratio. The optimization techniques are specifically tailored to ensure a zero dropped task ratio for urgent UEs' tasks and prioritize processing for them, while also aiming to minimize latency for all UEs. This methodology offers a more comprehensive evaluation of system efficiency. The main contributions of this paper are summarized as follows:

\begin{enumerate}
    \item [1)] This approach specifically targets the offloading of computationally intensive tasks. We examine the impact of networking parameters, including the quantity of UEs and tasks, on the latency and dropped task ratio. Furthermore, the impact of these factors on various scheduling techniques,including deterministic, heuristic, and optimization approaches, has been examined.
    \item [2)] An optimization problem is formulated based on our particular case, with the objective of minimizing both communication and computational latency and the dropped task ratio. Based on a thorough analysis of the literature, this work uniquely considers the combined factors of dropped task ratio, latency, and task urgency in a 5G-MEC scenario. Three optimization methodologies are employed, namely Mixed Integer Linear Programming (MILP), Particle Swarm Optimization (PSO), and Genetic Algorithm (GA). MILP is used as the primary optimization approach, and the results are compared with PSO and GA, which are reproduced based on our scenario, to assess its efficiency. The optimization results are compared with two scheduling algorithms, First Come First Serve (FCFS) and Shortest Task First (STF).
    \item[3)] We guarantee a zero dropped task ratio for urgent tasks in all optimization approaches by applying penalty in the objective function.
\end{enumerate}

The latency is decreased by almost 55\% compared to GA \cite{zhu21} and 35\% compared to PSO \cite{ma21} using the proposed MILP-based method. The proposed MILP-based technique reduces the dropped task ratio by around 70\% compared to GA \cite{zhu21} and by 40\% compared to PSO \cite{ma21}.


\begin{table*}[!t]
\caption{Comparison of our work with literature}
\centering
\renewcommand{\arraystretch}{1.8} 
\newcolumntype{M}[1]{>{\centering\arraybackslash}m{#1}}
\begin{tabular}{|M{1cm}|l|l|c|l|l|c|c|c|c|M{3cm}|}
\hline
\textbf{Paper} & \multicolumn{2}{c|}{\textbf{UE}} & \textbf{Tasks/UE} & \multicolumn{2}{c|}{\textbf{Task}} & \textbf{Urgency} & \multicolumn{3}{c|}{\textbf{Metric}} & \textbf{Algorithm} \\ \cline{2-3} \cline{5-6} \cline{8-10}
               & \textbf{Min} & \textbf{Max} & & \textbf{Min} & \textbf{Max} & & \textbf{Computational} & \textbf{Communication} & \textbf{Dropped Task} &                                 \\
               &              &              & &              & \multicolumn{1}{c|}{} & & \textbf{latency}       & \textbf{latency}         & \textbf{Ratio}       &                                 \\ \hline

\cite{farooq21}  & - & - & - &100 & 100 & \multicolumn{1}{c|}{\cmark} & \multicolumn{1}{c|}{\cmark} & \multicolumn{1}{c|}{\xmark} & \multicolumn{1}{c|}{\cmark} & nonlinear approach \\ \hline
\cite{fard24}  & 10 & 1000 & 1,2,5 &10 & 5000 & \multicolumn{1}{c|}{\xmark} & \multicolumn{1}{c|}{\cmark} & \multicolumn{1}{c|}{\xmark} & \multicolumn{1}{c|}{\cmark} & MILP, GA \\ \hline
\cite{li21}      & 50 & 250 & -  & 10 & 40 & \multicolumn{1}{c|}{\xmark} & \multicolumn{1}{c|}{\cmark} & \multicolumn{1}{c|}{\cmark} & \multicolumn{1}{c|}{\xmark} & Improved PSO \\ \hline
\cite{hsu22}     & 2500 & 2500 &- & - & - & \multicolumn{1}{c|}{\xmark} & \multicolumn{1}{c|}{\xmark} & \multicolumn{1}{c|}{\cmark} & \multicolumn{1}{c|}{\cmark} & DQN \\ \hline
\cite{zhu21}     & 2 & 20 &- & 1 & 200 & \multicolumn{1}{c|}{\xmark} & \multicolumn{1}{c|}{\cmark} & \multicolumn{1}{c|}{\cmark} & \multicolumn{1}{c|}{\xmark} & Improved GA \\ \hline
\cite{ma21}      & 1 & 500 & 1 & 1 & 500 & \multicolumn{1}{c|}{\xmark} & \multicolumn{1}{c|}{\cmark} & \multicolumn{1}{c|}{\xmark} & \multicolumn{1}{c|}{\xmark} & PSO \\ \hline
\cite{zeng22}    & - & - & - & - & - & \multicolumn{1}{c|}{\xmark} & \multicolumn{1}{c|}{\xmark} & \multicolumn{1}{c|}{\cmark} & \multicolumn{1}{c|}{\xmark} & ISSA, PSO \\ \hline
\cite{Seah2022}  & - & - & - &- & - & \multicolumn{1}{c|}{\xmark} & \multicolumn{1}{c|}{\xmark} & \multicolumn{1}{c|}{\cmark} & \multicolumn{1}{c|}{\xmark} & EWFQ/LC \\ \hline
\cite{liu22}     & 10 & 50 & 1 & 10 & 50 & \multicolumn{1}{c|}{\xmark} & \multicolumn{1}{c|}{\cmark} & \multicolumn{1}{c|}{\cmark} & \multicolumn{1}{c|}{\xmark} & nonlinear approach \\ \hline

\cite{feng21}    & 60 & 120 & 1 &60 & 120 & \multicolumn{1}{c|}{\xmark} & \multicolumn{1}{c|}{\cmark} & \multicolumn{1}{c|}{\xmark} & \multicolumn{1}{c|}{\xmark} & WOA+GWO \\ \hline
\cite{cheng22}   & 10 & 20 & 1 & 10 & 20 & \multicolumn{1}{c|}{\xmark} & \multicolumn{1}{c|}{\cmark} & \multicolumn{1}{c|}{\cmark} & \multicolumn{1}{c|}{\xmark} & D3QN \\ \hline 
\cite{alam19}   & 5 & 50 & 5 & 50 & 250 & \multicolumn{1}{c|}{\xmark} & \multicolumn{1}{c|}{\cmark} & \multicolumn{1}{c|}{\xmark} & \multicolumn{1}{c|}{\xmark} & MIP \\ \hline 
\cite{fengm21}   & - & - & - & - & - & \multicolumn{1}{c|}{\xmark} & \multicolumn{1}{c|}{\cmark} & \multicolumn{1}{c|}{\xmark} & \multicolumn{1}{c|}{\xmark} & MIP \\ \hline
\cite{gao19}   & 4 & 4 & 1 & 4 & 4 & \multicolumn{1}{c|}{\xmark} & \multicolumn{1}{c|}{\cmark} & \multicolumn{1}{c|}{\xmark} & \multicolumn{1}{c|}{\xmark} & MILP \\ \hline
\noteblue{\cite{apos2020}}   & - & 50 & - & - & - & \multicolumn{1}{c|}{\xmark} & \multicolumn{1}{c|}{\cmark} & \multicolumn{1}{c|}{\xmark} & \multicolumn{1}{c|}{\xmark} & Game theory+proposed algorithm \\ \hline
\noteblue{\cite{Li2023}}  & - & - & - & 3000 & 7000 & \multicolumn{1}{c|}{\xmark} & \multicolumn{1}{c|}{\cmark} & \multicolumn{1}{c|}{\cmark} & \multicolumn{1}{c|}{\xmark} & Hybrid GA-PSO \\ \hline
Ours             & 10 & 1000 &1,2,4 & 10 & 4000 & \multicolumn{1}{c|}{\cmark} & \multicolumn{1}{c|}{\cmark} & \multicolumn{1}{c|}{\cmark} & \multicolumn{1}{c|}{\cmark} & MILP, PSO, GA \\ \hline
\end{tabular}

\label{tab:gap}
\end{table*}

\section{Related Work}
Efficient task offloading can not only reduce energy consumption but also quicken task execution and enhance the UE experience \cite{akhlaqi23}. Computational offloading has been widely investigated in MEC networks, with the requirement that only one edge server handle all UE requests \cite{feng22}\cite{zeng22}. In \cite{zeng22}, Zeng et al. tackle the optimization issue that jointly optimizes the offloading rate, the proportion of the tasks that are compressed during offloading, the power used for transmission, and the allocation of CPU resources in the MEC server. The cost is formulated in terms of these parameters, and the objective is to minimize the overall cost of the system. In order to address the issue, the implementation of an improved sparrow search algorithm (ISSA) is employed.

In \cite{farooq21}, two approaches based on priority are introduced.
to allocate CPU cycles to offloaded tasks at an edge server.
In the initial approach the edge server is involved in discriminating
between various tasks by their priority classes. A greater number of CPU cycles per unit of time is assigned by the edge server to tasks in a higher priority class. In the second approach, the priority of each device is determined in real-time by considering the priorities of the tasks that the device has delegated. The edge server's CPU resources are assigned to a task according to the priority of the device that has delegated the task to the edge server. They improve the average computational latency of 100 UEs by 50\% compared with the round robin scheduling algorithm. However, they didn't consider communication latency in their work. Optimizing both communication and computational latencies ensures efficient data transfer and processing, particularly for latency-sensitive applications. In our scenario for 100 tasks, the latencies of urgent tasks are improved in the proposed MILP-based, PSO, and GA by an average of 77\%, 68\%, and 55\% compared to FCFS.

In addition, there has been a greater emphasis on computational offloading in multi-server MEC networks. This presents a greater challenge since the offloading decision involves not only determining whether a request should be offloaded or not but also determining the most suitable location for the offloading process. In \cite{Seah2022}, Seah et al. examine the impact of joint resource allocation and offloading in MEC. To meet the requirements of quality of service, this approach takes into account the communication and computational requirements of packets. A combined scheduling method called Extended Weighted Fair Queueing with Latency Constraint (EWFQ/LC) is proposed. This methodology tries to optimize fairness in resource allocation while also meeting the latency requirements of various service types. When there are sufficient resources, the scheduler functions similarly to Weighted Fair Queueing (WFQ) in order to equitably allocate both communication and compute resources.

In a multi-MEC server environment, users may prioritize risk-seeking or loss-aversion when deciding which computing tasks to offload to each server and execute locally. A probability of failure function is established for each MEC server, indicating its likelihood of failing to deliver end-user computing demands due to over-exploitation of its capabilities \cite{apos2020}. To enhance perceived satisfaction, authors in \cite{apos2020}, consider that users  offload computational tasks to several MEC servers using a prospect-theoretic utility function. To find the best allocation strategy, a non-cooperative game among users is proposed to determine the Pure Nash Equilibrium (PNE) for optimal data offloading.

Dynamic caching in MEC is a process that involves caching and managing data or information in a manner that adjusts to varying situations, UE requirements, or network characteristics. Dynamic caching differs from static caching in that it continuously updates and replaces cached content in response to real-time situations, as opposed to preloading and maintaining unchanged content until manually updated. The authors in \cite{liu22} examine a collaborative task offloading problem in MEC that is supported by dynamic caching. The objective is to minimize the maximum total latency experienced by all UEs. This is achieved by simultaneously evaluating the decisions related to service caching, task offloading, and computational resource allocation. A fair allocation of computer resources is proposed to enhance equity among UEs, where fairness is attained by minimizing the UEs' maximum actual latency. Offloading multi-dependent tasks in MEC presents a distinct and significant challenge within the broader context of task offloading. Multi-dependent tasks are a set of computational tasks that contain interactions which must be executed in a specified order. These tasks are essential parts of a broader computational workload or application and necessitate synchronized execution, wherein the output of one task serves as the input for another. Ma et al. \cite{ma21} propose a solution for multi-dependent tasks that involves utilizing a queue-based improved multi-objective particle swarm optimization algorithm. They take into account both the time it takes to complete the tasks and the cost of executing them. 

Feng et al. \cite{feng21} examine a computational offloading model that incorporates latency optimization, energy optimization, and price optimization for computational offloading in MEC.
They examine a model that takes into account these objectives and suggest a combination of the whale optimization algorithm (WOA) with the grey wolf optimizer (GWO). They apply the social hierarchy from GWO to WOA to reach global optimal while avoiding local optimal. The procedure for improving search agent positions is adjusted to accommodate the introduction of hierarchy. The encircling stage of WOA is similar to that of GWO, however, the bubble-net attack approach differs from GWO's hunting stage.

Li et al. \cite{Li2023}  propose a multi-objective optimization problem to reduce transmission delays caused by system load balancing and superfluous hops. They present GA-PSO, a hybrid genetic particle swarm technique. Their simulation results indicate that the hybrid GA-PSO method does not outperform cutting-edge GA and nearest selection algorithms in achieving all task offloading delays.

Zhu et al.\cite{zhu21} analyze how offloading decisions, uplink power allocation, and computing resource allocation affect latency.The objective function is defined as the weighted sum of task execution latency and energy usage, which is solved by improved GA.
Li et al. \cite{li21} aim to reduce the latency and energy usage resulting from the implementation of the MEC server in a 5G communication setting. An enhanced PSO algorithm is offered as an approach for offloading. The model aims to minimize communication latency while considering energy consumption limitations. To achieve a trade-off between latency and energy consumption, a penalty function is incorporated. The suggested computational offloading choice effectively assigns the calculation task to the appropriate MEC server. The Dynamic Task Offloading and Scheduling problem (DTOS) is formulated as a Mixed Integer Programming (MIP) called DTOS-MIP in \cite{alam19}. They explore the Logic Based Benders Decomposition (LBBD) technique as a means to effectively solve the DTOS problem and achieve optimal outcomes. The DTOS-LBBD algorithm partitions the DTOS problem into a primary task that handles task offloading and resource allocation, while also breaking it down into multiple sub-problems, each concentrating on the scheduling of tasks for one of the five applications.

Task partitioning is an alternate approach that can be used to decrease latency in MEC \cite{islam21}. If a computational task can be divided into smaller subtasks and assigned to various entities for execution, including the local device, MEC server, and cloud server, the workload at each entity can be decreased \cite{islam21}. Feng et al. \cite{fengm21} investigate the problem of task partitioning and UE association in a MEC system. The objective is to minimize the average latency per UE. It is presumed that every task can be divided into several subtasks that can be performed on local devices (such as cars), MEC servers, and/or cloud servers. The subtasks are either independent or dependent. Every UE is linked to one of the nearby edge nodes. The problem of jointly optimizing task partitioning and UE association is formulated as a MIP problem. For subtask dependency, they break down the problem into two separate subproblems. The first subproblem aims to optimize the ratio of task partitioning under a specific UE association, while the other subproblem is the association of UEs itself. The partitioning is pertained by the lower-level subproblem, whereas the association of UEs is involved by the higher-level subproblem.

Due to the increased computational complexity and data volume associated with Deep Neural Network (DNN) tasks compared to classical tasks, Mingjin et al. \cite{gao19} provide a viable solution by employing task partitioning and offloading.
They examine a DNN-based MEC system, taking into account several mobile devices and a single MEC server. In order to improve the task partitioning latency, they initially propose an approach that predicts the processing latency for DNN tasks. The goal is to minimize processing time and the computational load on mobile devices. A MILP-based mechanism is proposed to partition and offload DNN tasks.

Deep Reinforcement Learning (DRL) is widely used by researchers for task offloading in MEC \cite{zabihi23}. DRL is particularly effective in dealing with non-linearities and multi-objective optimization in complex systems, where conventional approaches may encounter difficulties or require linearization \cite{hort23}. It effectively achieves a harmonious equilibrium between several goals, such as efficiency and speed, without necessitating complex mathematical models \cite{hua23}.
Yun et al. \cite{yun22} present a control scheme based on DRL and GA to meet the demanding QoS requirements using 5G multiple radio access technology (RAT)-based partial offloading and MEC resource allocation. In the proposed approach, the MEC server dynamically changes its resources depending on offloading requests from different UEs utilizing DRL. The GA decides whether a task should be processed locally or offloaded. The objective of the proposed approach is to reduce energy usage and increase the offloading success rate of servers.

Hsu et al. \cite{hsu22} study computational offloading in a heterogeneously loaded MEC-enabled Internet of Things (IoT) network. A heterogeneous workload regarding MEC refers to a scenario where the computational tasks or workloads show significant variations in terms of their resource requirements, complexity, and characteristics. In other words, tasks and UEs vary in their computational requirements and processing needs. This work models the computational offloading as a Markov decision problem and develops a Duelling-Deep Q-learning Network (DQN)-based energy and latency-efficient technique.
Cheng et al. \cite{cheng22} focus on investigating the methods and techniques used for offloading and allocating resources in MEC with the goal of minimizing latency and energy. Based on the characteristics of the objective function, a Double-Duelling DQN (D3QN) algorithm is suggested to address the issues of overfitting and overestimation in the DQN algorithm.

We summarize all these works based on multiple factors, as shown in Table~\ref{tab:gap}.
Based on our thorough review of the available literature, we have discovered an important gap in current research regarding the incorporation of the dropped task ratio with other network parameters, especially computational latency. In our previous work in \cite{fard24}, we consider dropped task ratio and computational latency and ignore communication latency. Also, in \cite{hsu22}, they consider minimizing dropped task ratio with communication latency, ignoring computational latency. In this work, computational latency, communication latency, and dropped task ratio are considered. Furthermore, a limited number of articles take into account the urgency of tasks in their optimization problems. In \cite{farooq21}, they consider task urgency, but with a limited number of tasks and an unknown number of UEs. In this paper, we have up to 4000 tasks and 1000 UEs and aim to minimize not only dropped task ratio and computational latency but also communication latency. This article aims to bridge the current gap by: 1) reducing communication and computational latency, and 2) minimizing the dropped task ratio.

\begin{figure}[!hbt]
        \centering
        \includegraphics[width = 0.5\textwidth, trim=0cm 0cm 0cm 0cm,clip]{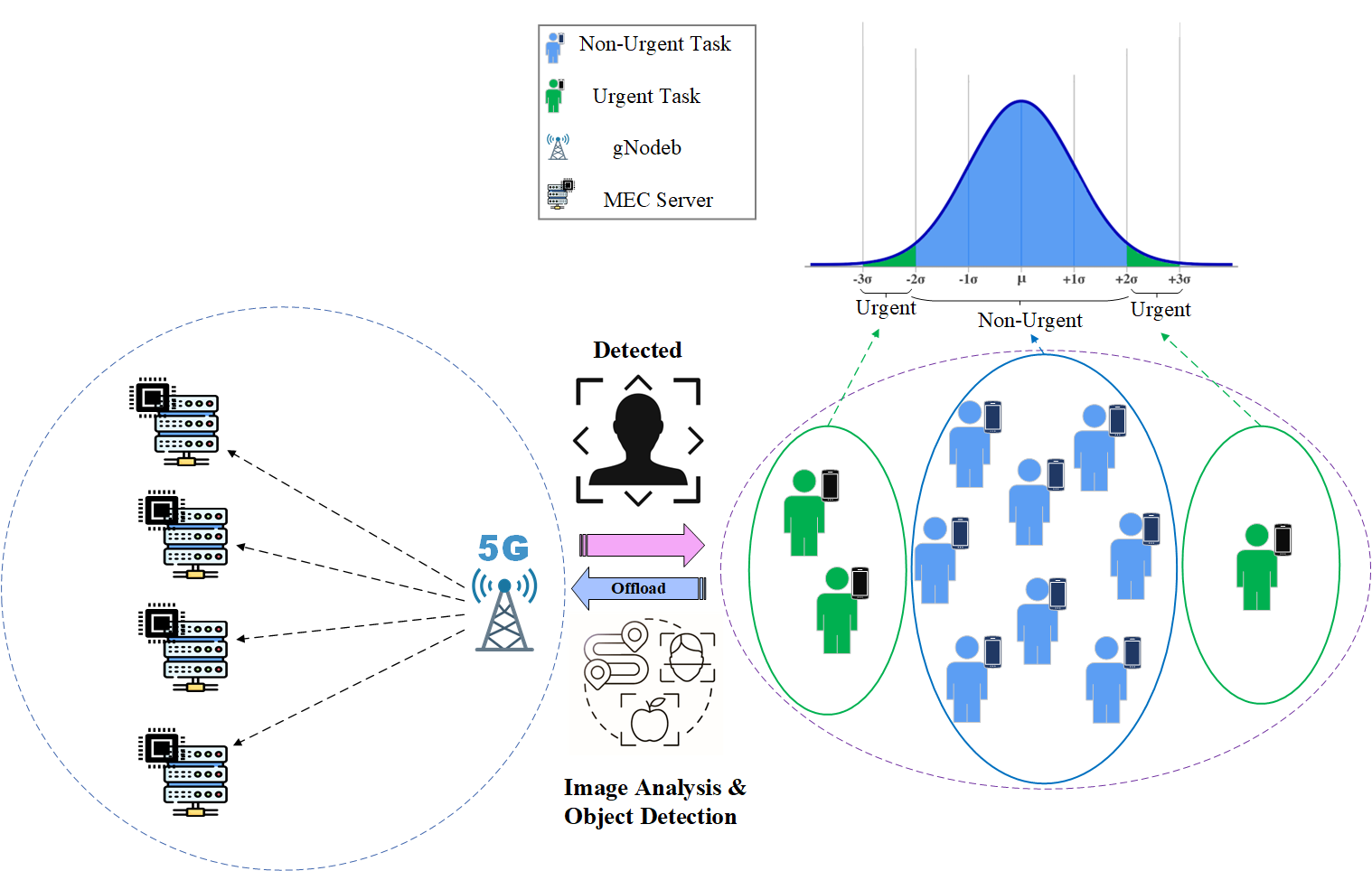}
        \caption{ Proposed methodology}
        \label{fig:methodd} \vspace{-2mm}
\end{figure}

\begin{figure*}[!hbt]
        \centering
        \includegraphics[width = 0.7\textwidth, trim=1.3cm 3cm 1.5cm 1.5cm,clip]{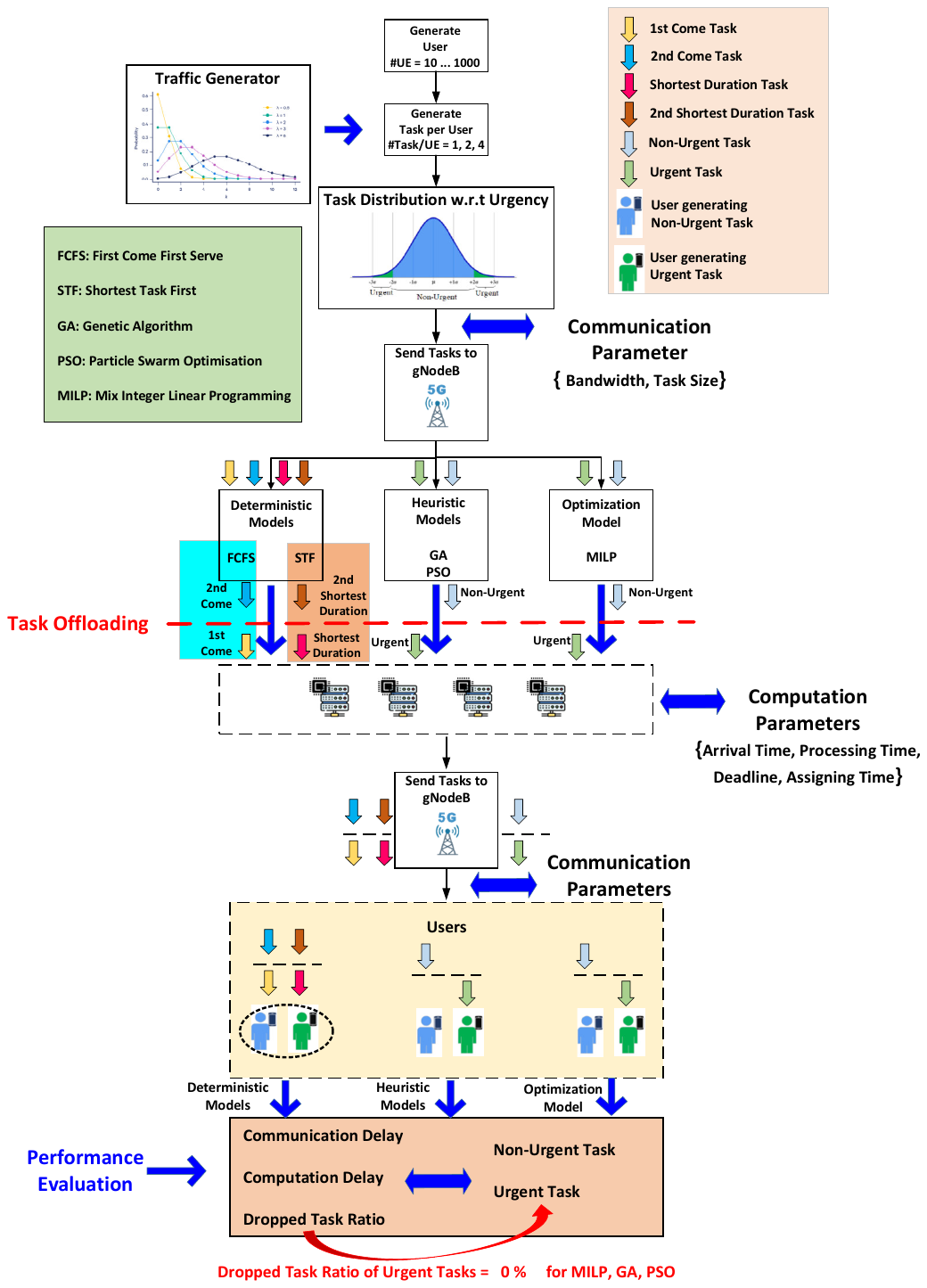}
        \caption{General Process Flow for Proposed Methodology}
        \label{fig:flowchart} \vspace{-2mm}
\end{figure*}


\section{Methodologies under Study} 

We consider a multi-UE multi-server 5G-based MEC, as shown in 
\figurename\ref{fig:methodd}. The MEC servers are deployed near a gNB to handle
tasks offloaded from UEs and send back the results to them. Urgent and non-urgent tasks have been classified in order of priority in this study.

The \figurename\ref{fig:flowchart} illustrates a flowchart for the proposed methodology, focusing on how to manage task allocation based on resource availability to minimize latency and drop task ratio. It starts with task generation with poisson distribution and then sends the tasks to gNodeB and MEC servers. Tasks are categorized as either urgent or non-urgent based on a Gaussian distribution, where from -3std to -2std and 
 from +2std to +3std are classified as urgent tasks and the rest are classified as non-urgent tasks. Tasks are processed using different strategies, including two deterministic models, FCFS and STF, two heuristic models, GA and PSO, and one optimization model, MILP. The optimization models aim to optimize task offloading based on communication and computational parameters, leading to a performance evaluation that considers communication latency, computational latency, and the dropped task ratio. The goal is to efficiently manage task offloading while minimizing total latency and dropped task ratio, particularly ensuring that urgent tasks have a zero percent dropped task ratio and are processed first.

\subsection{Greedy Schemes} \textbf{FCFS} assigns tasks to computing resources in the order they arrive, without considering their computational requirements or deadlines. Each incoming task is placed in a queue and waits for its turn to be processed. This offloading strategy, although easy to implement and fair in terms of task execution order, can lead to high waiting times, especially for lightweight tasks that are queued behind heavier ones \cite{Ali2021}.

\textbf{STF} is a scheduling method that prioritizes tasks with shorter processing times over those with longer processing times. The main goal of this strategy is to minimize the total execution time while also targeting the task completion ratio. By prioritizing shorter tasks, it reduces latency for varying task sizes in scalable environments. Although FCFS is fair and straightforward in task queue management, its sequential processing might increase latency, especially with large tasks upfront, making it less scalable compared to STF. This strategy has the potential to significantly reduce the average task waiting time. However, there is a possible disadvantage to this technique as it increases the likelihood of longer tasks being kept in a queue, as they may consistently be assigned lower priority levels \cite{Yin2022}. The feasibility of implementing the STF in real-time systems is greatly impeded by its dependence on having prior knowledge about each task, a condition that is frequently impractical or impossible in environments where task sizes and characteristics can be unpredictable, a situation that is often impracticable or unattainable in real-time systems where task sizes can be unpredictable.

 \begin{table}[!hbt]
\centering
\renewcommand{\arraystretch}{1.4} 

\caption{Notation table for parameters used in the optimization problem.}
\label{tab:notation}
\begin{tabularx}
{0.4\textwidth}{|c|X|}
\hline
\textbf{Parameter} & \textbf{Decription} \\
\hline
\( N \) & Set of tasks \\
\hline
\( M \) & Set of servers \\
\hline
\( B_i \) & Bandwidth of channel for the UE of task \(i\) \\
\hline
\( p \) & Transmission power of UE \\
\hline
\( g \) & Channel gain \\
\hline
\( N_0 \) & Density of noise power \\
\hline
\( t_{d_i} \) & Deadline of task \( i \) \\
\hline
\( t_{a_i} \) & Arrival time of task \( i \) \\
\hline
\( t_{c_i} \) & processing time of task \( i \) \\
\hline
\( t_{s_i} \) & Start processing time of task \( i \) \\
\hline
\( x_{ij} \) & Binary decision variable which is 1 when task i is assigned to server j.\\
\hline
\( f_{j} \) & Frequency of cpu on server j\\
\hline
\( S_i \) & Size of the task\\
\hline
\( C_i \) & Cycles needed to complete the task i\\
\hline
\( M_{itj} \) & Integer scheduling matrix for task i scheduled on server j in timeslot t.\\
\hline
\end{tabularx}

\end{table} 

\subsection{Mathematical foundations of task offloading}

The optimization problem involves an objective function that seeks to minimize both latency and dropped task ratio. The goal of objective function is a trade-off between the two baseline methods, FCFS and STF. 
Here, we aim to minimize both latency and dropped task ratio while considering deadline, duration, urgency and task arrival time. 

\subsubsection{computational}
We  use  a binary decision variable, \( x_{ij} \), for assigning tasks to servers where:
\vspace{-1mm}
\begin{equation}
\hspace{-8mm}
  x_{ij} = 
  \begin{cases} 
    1 & \text{ task } i \text{ is assigned to server } j  \\
    0 & \text{otherwise}
  \end{cases}
  \label{eq:two}
\end{equation}

We formulate computational latency as in (\ref{eq:complatency}), waiting time, and (\ref{eq:comexecute}), execution latency. In (\ref{eq:complatency}), \( N \) is the set of tasks and \( M \) is the set of servers, \( t_{s_i} \) is the start processing time of task \( i \), \( t_{a_i} \) is the arrival time of task \( i \), \( t_{d_i} \) is the deadline of task \( i \) and \( t_{c_i} \) is the processing time of task \( i \). All of the required notation for mathematical formulas is briefly explained in \tablename \hspace{0.1pt} \ref{tab:notation}.

\begin{equation}
\label{eq:complatency}
T_{i}^w= \frac{t_{s_i} - t_{a_i}}{t_{d_i} - t_{a_i} - t_{c_i}}
\end{equation}

The execution latency for task i on server j is formulated as  (\ref{eq:comexecute}), where \(f_{j}\) is the frequency of CPU in server j and \(C_{i}\) is the cycles needed to complete task i.
\begin{equation}
T_{ij}^e = \frac{C_i}{f_{j}}
\label{eq:comexecute}
\end{equation}

We formulated, \(T_{ij}^c\), the computational latency as (\ref{eq:comp}):
\begin{equation}
T_{ij}^c = T_{ij}^e + T_{i}^w
\label{eq:comp}
\end{equation}

\subsubsection{Communication latency}
the uplink data rate \cite{elgendy20} for the UE of task i is given as:

\begin{equation}
r_{i} = B_{i} \log_2 \left(1 + \frac{p g}{N_0}\right)
\label{eq:datarate}
\end{equation}

where \(B_i\) denotes the bandwidth of the channel for the UE of task i, \(p\) is the transmission power of the UEs that is identical for all, \(g\) represents the gain of the channel, and \(N_0\) is the density of the noise power of the channel. To manage medium access and mitigate interference between UEs\cite{tan21}, Orthogonal Frequency-Division Multiple Access (OFDMA) is employed \cite{huo24}. The transmission latency can be expressed as formulated in  (\ref{eq:comlatency}), where \(S_i\) is size of the task \(i\). We consider roundtrip communication latency in this paper.

\begin{equation}
T_{i}^t = 2 \times \frac{S_i}{r_i}
\label{eq:comlatency}
\end{equation}

\subsubsection{Dropped Task Ratio}

The dropped task ratio is formulated in (\ref{eq:dtr}):

\begin{equation}
D_{ij} = \frac{1 - x_{ij}}{N}
\label{eq:dtr}
\end{equation}

\subsubsection{The optimization formula}
Our goal is to minimize total latency (communication + computational) for all offloaded tasks along with the dropped task ratio as in (\ref{eq:obj1}). Thus, tasks are offloaded with the lowest latency, considering minimizing the dropped task ratio. Hence, the objective function is formulated as in (\ref{eq:obj2}), a NP-hard problem, on the basis of (\ref{eq:c1})-(\ref{eq:Mitj}), which will be linearized for MILP algorithm in part 5. The problem is NP-hard, since it's nonlinear and non-convex and becomes more and more complex as the number of tasks increases \cite{chen2021}. 
The binary nature of \(x_{ij}\) implies that the feasible set is not a convex set. Convexity is typically considered over continuous domains. When it is restricted to binary values, the function can exhibit non-convex behavior due to the discrete nature of the domain.

In our system, urgent tasks are prioritized, ensuring that they are processed earlier than non-urgent tasks. A penalty ,\(\delta\), is incorporated to address the potential consequences of prioritizing urgent tasks over the execution of non-urgent tasks. The penalty serves the purpose of maintaining balance in the system's operations, guaranteeing that urgent tasks are first processed while also maintaining efficiency in the overall processing of tasks.

The objective function is subject to the constraints in (\ref{eq:c1})-(\ref{eq:Mitj}). The constraint in (\ref{eq:c1}) ensures that each task is assigned to at most one server. In (\ref{eq:bandwidth}), an upper bound of $B_{max}$ is set on the bandwidth. In (\ref{eq:tsp1}), the first task will start being processed upon its arrival. In (\ref{eq:tsp}), the start processing time of a task is between its arrival time and task deadline, taking into account processing and communication latency. It ensures that the task begins processing at a time that permits it to be completed before the deadline, taking into account communication latency. In (\ref{eq:Mitj}) the total time spent on task \(i\) when scheduled on server \(j\) during time slot \(t\) does not exceed the processing time of the task. We define \(M_{itj}\) as an integer scheduling matrix for task i on server j in timeslot t.

\begin{equation}
     \sum_{j=1}^{M} \sum_{i=1}^{N} \delta(T_{i}^t + T_{ij}^c) \times x_{ij}  + \sum_{j=1}^{M} \sum_{i=1}^{N} D_{ij}
    \label{eq:obj1}
\end{equation}
\begin{equation}
\begin{aligned}
    & \sum_{j=1}^{M} \sum_{i=1}^{N} \delta \left(  \frac{t_{\text{s}_i} - t_{a_i}}{t_{d_i} - t_{a_i} - t_{c_i}}  +\frac{C_i}{f_j} + 2 \times \frac  {S_i}{r_{i}}  \right) \times x_{ij} + \\
    & \sum_{j=1}^{M} \sum_{i=1}^{N} \frac{1 - x_{ij}}{N}
\end{aligned}
\label{eq:obj2}
\end{equation}

\begin{equation}
S.t. :     
\sum x_{ij} \leq 1   
\label{eq:c1}
\end{equation}

\begin{equation}
    B_{i} \leq B_{\text{max}}
    \label{eq:bandwidth}
\end{equation}
\begin{equation}
    t_{\text{s}_1} = t_{a_1}
    \label{eq:tsp1}
\end{equation}
\begin{equation}
    t_{a_i} \leq t_{\text{s}_i} \leq t_{d_i} - t_{c_i} -  T_{i}^t
    \label{eq:tsp}
\end{equation}

\begin{equation}
    \sum_{j} \sum_{t} x_{ij} \cdot M_{itj} \leq t_{c_i}
    \label{eq:Mitj}
\end{equation}

\subsubsection{Linearizing}
\label{linearize}

As the optimization problem is formulated in the form of an MILP model, we need to linearize nonlinear equations. Thus, given (\ref{eq:p1}):  

\begin{equation}
    x_{ij} \cdot A \geq 0
    \label{eq:p1}
\end{equation}

The provided set of inequalities defines a linearization process, commonly employed to transform nonlinear problems, \(B = x_{ij} \cdot A\), into a linear format that is more manageable to solve. This non-linear to linear transformation is performed according to (\ref{eq:l1}) to (\ref{eq:l3}). Here, \(A\) takes values from the set \{0,...,Max$_a$\}. 
\begin{align}
B - A &\leq 0 \label{eq:l1} \\
B - x_{ij} &\leq 0 \label{eq:l2} \\
x_{ij} + A - B &\leq Max(A)\label{eq:l3}
\end{align}
For instance, the (\ref{eq:Mitj}),  \(T_{itj} = x_{ij} \cdot M_{itj}\) is linearized as (\ref{eq:mitjl1}) to (\ref{eq:mitjl3}). Here, \(Q_{itj}\) is the maximum value of \(M_{itj}\).
\begin{align}
T_{itj} - x_{ij} &\leq 0 \label{eq:mitjl1} \\
T_{itj} - M_{itj} &\leq 0 \label{eq:mitjl2} \\
x_{ij} + M_{itj} - T_{itj} &\leq Q_{itj}\label{eq:mitjl3}
\end{align}
We apply the same process for \( (T_{i}^t + T_{ij}^c) \times x_{ij}\) in (\ref{eq:obj1}). If \( T_{i}^t  \times x_{ij} = T_o\), we can linearize it according to (\ref{eq:to1}) to (~\ref{eq:to3}):
\begin{align}
T_o - x_{ij} &\leq 0 \label{eq:to1} \\
T_o - T_{i}^t &\leq 0 \label{eq:to2} \\
x_{ij} + T_{i}^t - T_o &\leq Max(T_o)\label{eq:to3}
\end{align}
For \( T_{ij}^c  \times x_{ij} = T_c\) in (\ref{eq:obj1}), we  linearize according to (\ref{eq:tc1}) to ~\ref{eq:tc3}:
\begin{align}
T_c - x_{ij} &\leq 0 \label{eq:tc1} \\
T_c - T_{ij}^c &\leq 0 \label{eq:tc2} \\
x_{ij} + T_{ij}^c - T_c &\leq Max(T_c)\label{eq:tc3}
\end{align}

\subsection{The Algorithm}

\begin{algorithm}
\color{black}
\caption{Task Scheduling and Optimization}
\begin{algorithmic}[1]

\STATE \textbf{1. Generate Users and Tasks:}
\FOR{$P \in \{10, 100, 500, 1000\}$}
    \FOR{$p = 1$ to $P$}
        \STATE Generate Poisson-distributed arrival times for tasks $i$.
        \FOR{$i = 1$ to $N$}
            \STATE Generate task distribution $G$ using Gaussian.
            \STATE Classify task $i$ as urgent if $G \in [-3\sigma, -2\sigma] \cup [2\sigma, 3\sigma]$, else non-urgent.
        \ENDFOR
    \ENDFOR
\ENDFOR

\STATE \textbf{2. Deterministic Models:}

\STATE \textbf{2.1 FCFS:}
\FOR{$r = 1$ to $10$}
    \STATE Schedule tasks based on arrival times.
    \STATE Calculate $\mathscr{T}_f^o(r) = \sum (T_{i}^t \times x_{ij})$
    \STATE Calculate $\mathscr{T}_f^c(r) = \sum (T_{ij}^c \times x_{ij})$
    \STATE Calculate $\mathscr{D}_f(r) = \sum D_{ij}$
\ENDFOR
\STATE Get average of $\mathscr{T}_f^o(r)$, $\mathscr{T}_f^c(r)$, and $\mathscr{D}_f(r)$ on $r$

\STATE \textbf{2.2 STF:}
\FOR{$r = 1$ to $10$}
    \STATE Schedule tasks based on shortest durations.
    \STATE Calculate $\mathscr{T}_s^o(r) = \sum (T_{i}^t \times x_{ij})$
    \STATE Calculate $\mathscr{T}_s^c(r) = \sum (T_{ij}^c \times x_{ij})$
    \STATE Calculate $\mathscr{D}_s(r) = \sum D_{ij}$
\ENDFOR
\STATE Get average of $\mathscr{T}_s^o(r)$, $\mathscr{T}_s^c(r)$, and $\mathscr{D}_s(r)$ on $r$

\STATE \textbf{3. Handling Urgent Users:}
\FOR{$i \in N$}
    \IF{$i \in U$}
        \STATE Prioritize task $c$ with $D_{ij} = 0$.
    \ENDIF
\ENDFOR
\FOR{$i \in U$}
    \STATE Process urgent tasks first.
\ENDFOR

\STATE \textbf{4. Optimization:}

\STATE \textbf{MILP:}
\FOR{$r = 1$ to $10$}
    \STATE Minimize (9) based on (10) to (27).
    \FOR{$i \in U$}
        \STATE Calculate $\mathscr{T}_1^o(r) = \sum (T_{i}^t \times x_{ij})$
        \STATE Calculate $\mathscr{T}_1^c(r) = \sum (T_{ij}^c \times x_{ij})$
    \ENDFOR
    \FOR{$i \in U'$}
        \STATE Calculate $\mathscr{T}_2^o(r) = \sum (T_{i}^t \times x_{ij})$
        \STATE Calculate $\mathscr{T}_2^c(r) = \sum (T_{ij}^c \times x_{ij})$
        \STATE Calculate $\mathscr{D}_2(r) = \sum D_{ij}$
    \ENDFOR
\ENDFOR
\STATE Get average of $\mathscr{T}_1^o(r)$, $\mathscr{T}_1^c(r)$, $\mathscr{T}_2^o(r)$, $\mathscr{T}_2^c(r)$, and $\mathscr{D}_2(r)$ on $r$

\STATE \textbf{5. Result Compilation and Analysis:}
\STATE Compile results including task assignments sequences, total delays, and dropped task ratios.

\end{algorithmic}
\label{a1}
\end{algorithm}

The algorithm \ref{a1} is a comprehensive approach to task scheduling and optimization, considering both deterministic and optimization methods. It starts by generating \(P\) users and \(N\) tasks, where tasks arrive based on a Poisson distribution, and their urgency is determined using a Gaussian distribution.

In the deterministic models section, two scheduling approaches are employed. For FCFS, tasks are scheduled based on their arrival times. For each run \(r\) , the total offloading time \(\mathscr{T}_f^o(r)\), total computation time \(\mathscr{T}_f^c(r)\), and total dropped task ratio \(\mathscr{D}_f(r)\) are calculated and averaged over multiple runs. For STF, tasks are scheduled based on their shortest durations. Similar to FCFS, the metrics \(\mathscr{T}_s^o(r)\), \(\mathscr{T}_s^c(r)\), and \(\mathscr{D}_s(r)\) are calculated and averaged over multiple runs.
Urgent tasks are given special consideration. They are prioritized, and tasks classified as urgent (with a dropped task ratio of zero) are processed first.

We use MILP to solve the problem. The model aims to minimize a specified objective function (\ref{eq:obj2}) while adhering to constraints (10) to (27). For each run, Offloading and computation delay for both urgent \(U\) and non-urgent \(U'\) tasks are calculated, along with dropped task ratio for non-urgent tasks. These metrics are averaged across runs to assess the performance of the MILP approach. For GA and PSO, we re-simulate  \cite{zhu21} and \cite{ma21} proposed model.

\section{Performance Evaluation}
\subsection{Settings}
The simulations are carried out using a computer equipped with a core i7 CPU, 4080ti GPU, and 32GB of RAM. The scenario is simulated in MATLAB, and the parameters have been set according to \cite{Gao2023, Seah2022} as shown in Table~\ref{tab:parameters}. We simulate the 5G gNB with four MEC servers. On each MEC server, there is one CPU for processing tasks, and tasks follow a poisson arrival. The tasks are DL-based object detection based on YOLOv5 \cite{arda23}. Tasks are offloaded with respect to the FCFS and STF fashion. We formulate the objective function \ref{eq:obj2} and try to find the best offloading decision to minimize the dropped task ratio and latency. We compare our optimization results with baseline offloading schemes. The parameters used for GA \cite{zhu21} and PSO \cite{ma21} are shown in Table~\ref{tab:GA} and~\ref{tab:PSO}, respectively. These parameters are chosen empirically after a couple of experiments. The fitness function is defined as the objective function in (\ref{eq:obj2}).

\begin{table}[!hbt]
\centering
\renewcommand{\arraystretch}{1.3} 
\caption{Parameters values in simulations} 
\begin{tabular}{|c|c|}
\hline
\textbf{Parameters} & \textbf{Value}   \\   \hline
Number of UEs & 10,50,100,500,1000   \\   \hline
Number of tasks per UE& 1,2,4   \\   \hline
Max number of tasks & 4000   \\   \hline
Number of servers & 4   \\   \hline
Number of CPU in each server & 1   \\   \hline
CPU Speed &  2.2 Ghz   \\   \hline
Datarate & 50Mbps   \\   \hline
noise power & 100dbm   \\   \hline
transmit Power & 200 mW \\   \hline
\end{tabular}
\label{tab:parameters} \vspace{-1mm}
\end{table}

\begin{table}[!hbt]
\centering 
\renewcommand{\arraystretch}{1.3} 
\caption{Parameters for GA}
\label{table:ga_parameters}
\begin{tabular}{|l|l|}
\hline
\textbf{Parameter} & \textbf{Value/Type} \\
\hline
Population Size & 50 individuals in each generation \\
\hline
Max Generations & 200 \\
\hline
Mutation Rate & 0.01 \\
\hline
Mutation Type & @mutationgaussian \\
\hline
Crossover Type & @crossoverscattered\\
\hline
\end{tabular}
\label{tab:GA}
\end{table}
\begin{table}[!hbt]
\centering 
\renewcommand{\arraystretch}{1.2} 
\caption{Parameters for PSO}
\label{table:ga_parameters}
\begin{tabular}{|l|l|}
\hline
\textbf{Parameter} & \textbf{Value/Type} \\
\hline
Swarm Size & 50  \\
\hline
Cognitive Coefficient & 1.49 \\
\hline
Social Coefficient & 1.49 \\
\hline
Max Generation & 200 \\
\hline
\end{tabular}
\label{tab:PSO}
\end{table}

\begin{figure}[!hbt]
 \centering
    \begin{subfigure}{0.48\textwidth}
        \centering
        \includegraphics[width =\textwidth, trim= 4.5cm 0cm 4.2cm 0cm ,clip]{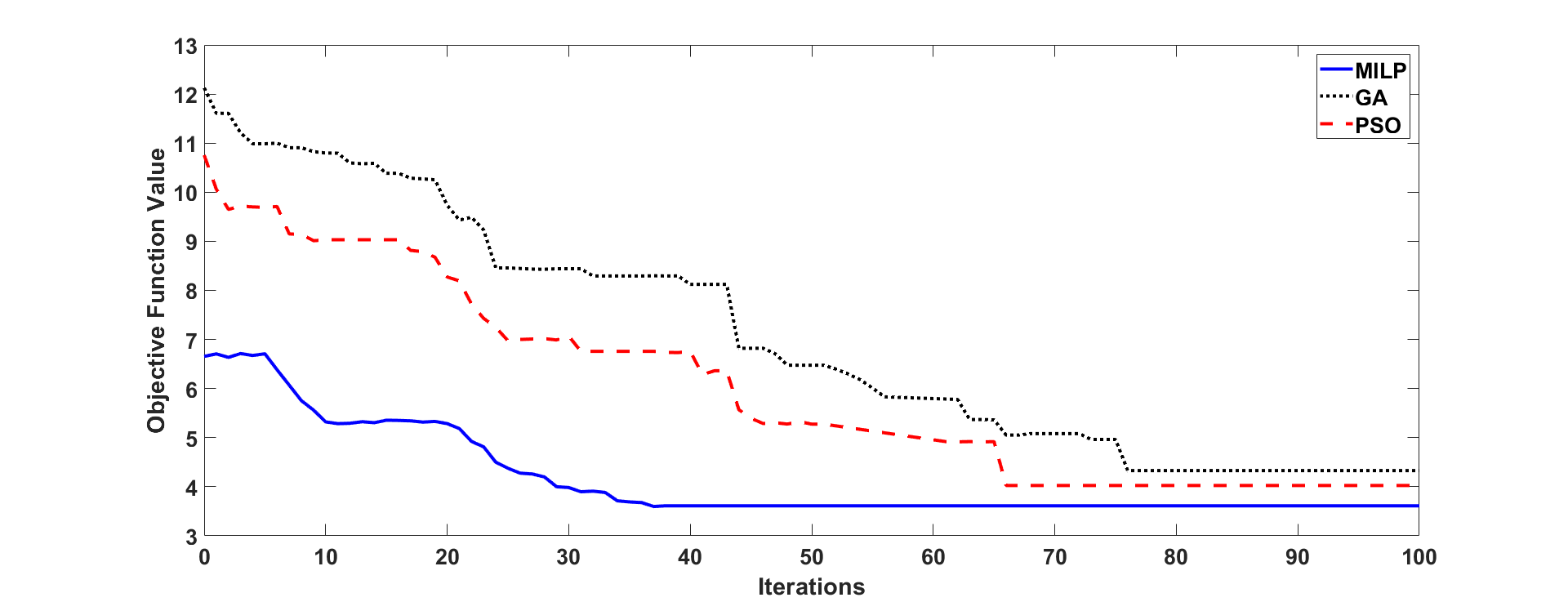}
        \caption{ One task per UE}
        \label{fig:convergence1}
    \end{subfigure}
    \begin{subfigure}{0.48\textwidth}
        \centering
        \includegraphics[width =\textwidth, trim= 4.5cm 0cm 4.2cm 0cm ,clip]{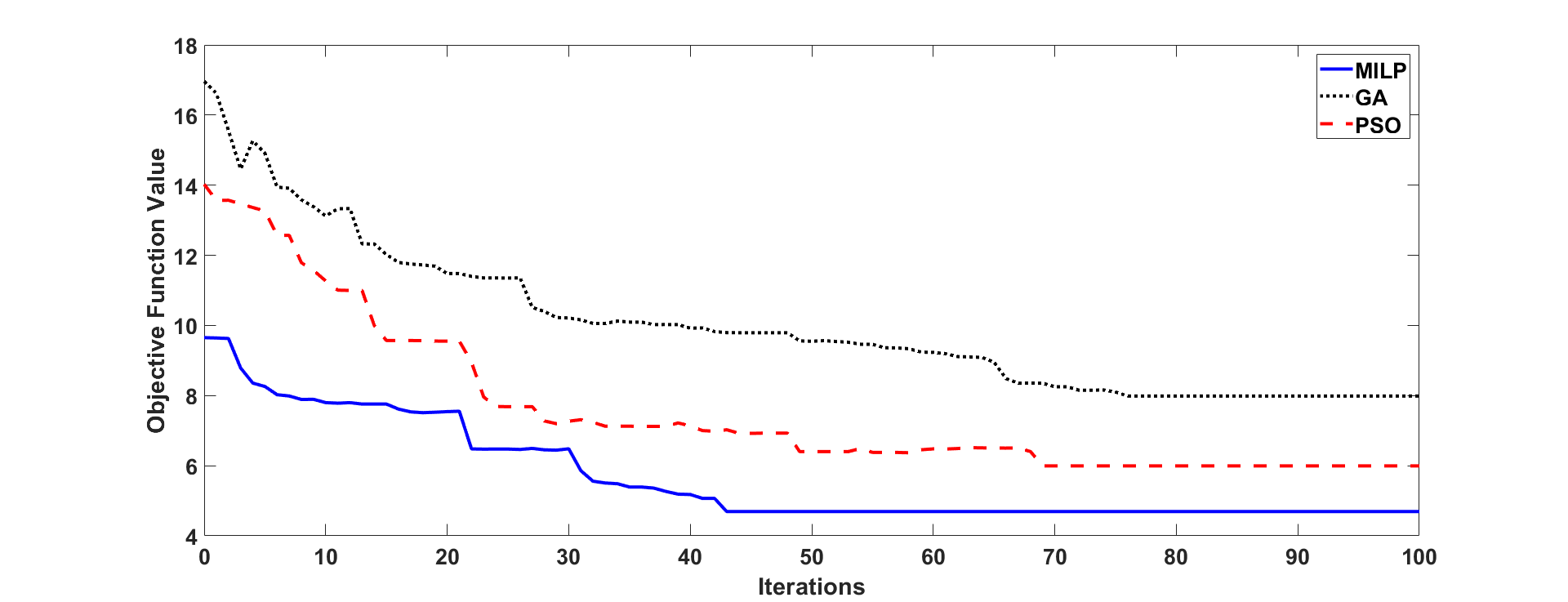}
        \caption{Two tasks per UE}
        \label{fig:convergence2}
    \end{subfigure}
    \begin{subfigure}{0.48\textwidth}
        \centering
        \includegraphics[width =\textwidth, trim= 4.5cm 0cm 4.2cm 0cm ,clip]{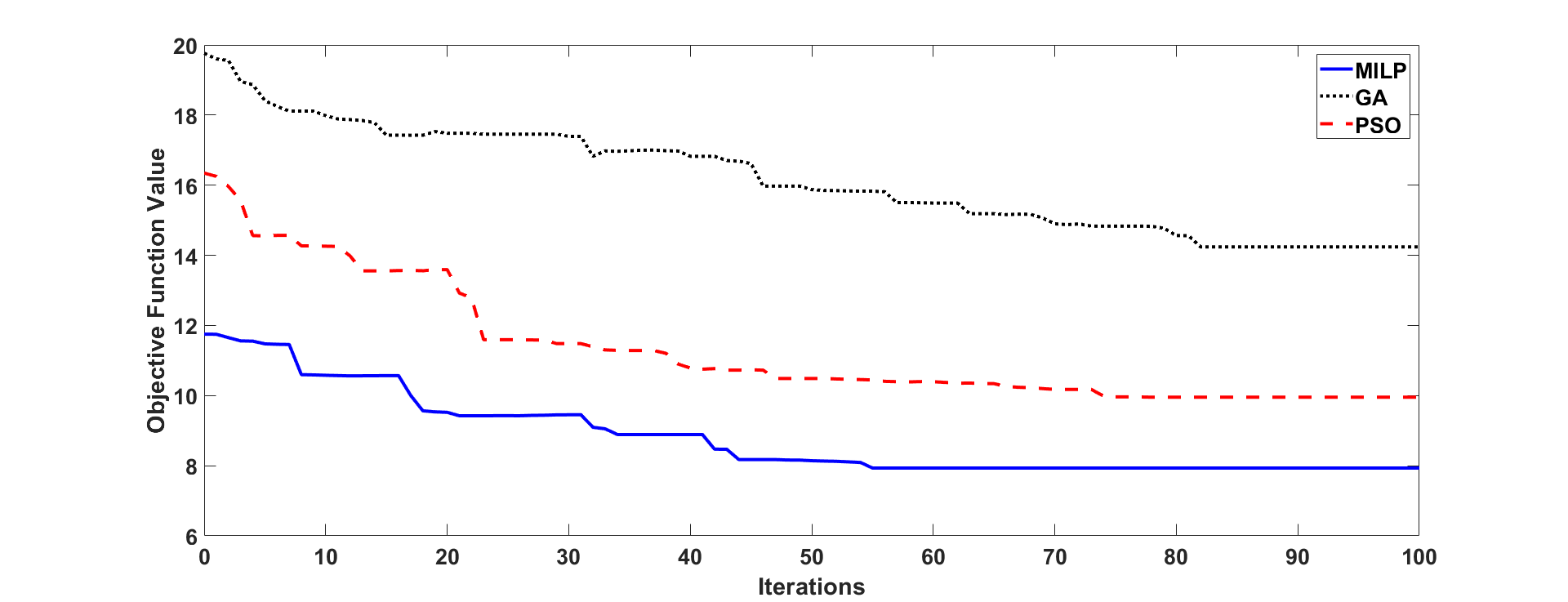}
        \caption{Four tasks per UE}
        \label{fig:convergence4}
    \end{subfigure}
        \caption{Convergence to the objective function's optimal value} 
	\label{fig:convergence} 
\end{figure}

\subsection{Experimental Results}
In this section, we present the convergence of algorithms and numerical results for communication latency, computational latency, and dropped task ratio for FCFS, STF, PSO, GA, and  MILP. We analyze how the simulated algorithms are affected by the quantity of UEs and tasks for each UE. The performance of each optimization strategy on both urgent and non-urgent tasks is analyzed as well.

\subsubsection{Convergence}
In \figurename \hspace{0.1pt} \ref{fig:convergence}, we plot the objective function values during iterations for different numbers of tasks per UE. 
In \figurename \hspace{0.1pt} \ref{fig:convergence1}, MILP shows efficient convergence towards an optimal solution, as seen by the rapid and consistent reduction in the value of the objective function in (\ref{eq:obj2}) across different tasks per UE. This trend demonstrates the algorithm's ability to rapidly identify and pursue the most desirable result, the lowest computational latency, and the lowest dropped task ratio. MILP can be used to minimize latency effectively if the offloading problem is defined as a set of linear equations and inequalities. The objective function value of the GA shows a gradual reduction, with rather steady plateaus interspersed with abrupt dips. This pattern is indicative of the GA's evolutionary search process, where a population of solutions is evolved over time. Improvements occur as a result of the selection, crossover, and mutation operators. The GA has progressive enhancements, but it does not achieve as minimal a value as the MILP-based solution, which makes it less efficient in our case. The convergence pattern of GA is identified by a slower decrease in the value of the objective function, which is associated with the increased computational latency and dropped task ratio for GA. It indicates that although GA may finally discover suitable solutions, but with lower computational efficiency, requiring more iterations to reach stability.

The PSO algorithm has a better objective function value than GA but underperforms when compared to MILP.In summary, the graph indicates that MILP outperforms the other algorithms according to not only the value of the objective function but also its faster convergence. Therefore, MILP has better efficiency for the optimization problem when taking into account the objective function in (\ref{eq:obj2}).

In \figurename \hspace{0.1pt} \ref{fig:convergence2}, the MILP-based solution constantly shows rapid and robust convergence, highlighting its efficacy in solving the given issue. Both GA and PSO show different levels of improvement, with PSO initially outperforming GA but ultimately underperforming when compared to MILP in terms of the objective function in this specific scenario.
In \figurename \hspace{0.1pt} \ref{fig:convergence4} MILP demonstrates the best performance by rapidly obtaining the lowest objective function values and consistently maintaining them, indicating robust and stable convergence towards an optimal solution. PSO starts with smaller initial objective function values than the GA, due to more effective particle positioning (solutions) in the search space or a better initial velocity distribution, allowing particles to explore promising regions early on \cite{gang18}. PSO fails to maintain or enhance its performance as iterations progress, resulting in a less optimal value than MILP. PSO relies on particle cooperation and information exchange, therefore, if particles converge early or get locked in local optima, continued development is difficult \cite{gang18}.
The GA improves slower and more inconsistently than MILP. GA uses random processes like mutation and crossover to develop new solutions, which causes this oscillation \cite{GA19}. These random processes can yield superior offspring that significantly decrease the objective function value or less fit solutions that temporarily increase or freeze it. GA's performance shows a lack of consistency, characterized by a broad pattern of improvement, albeit at a slower and more fluctuating pace compared to MILP.

\begin{figure}[!hbt]
 \centering
    \begin{subfigure}{0.5\textwidth}
        \centering
        \includegraphics[width =\textwidth, trim= 4.1cm 0cm 3.9cm 0cm ,clip]{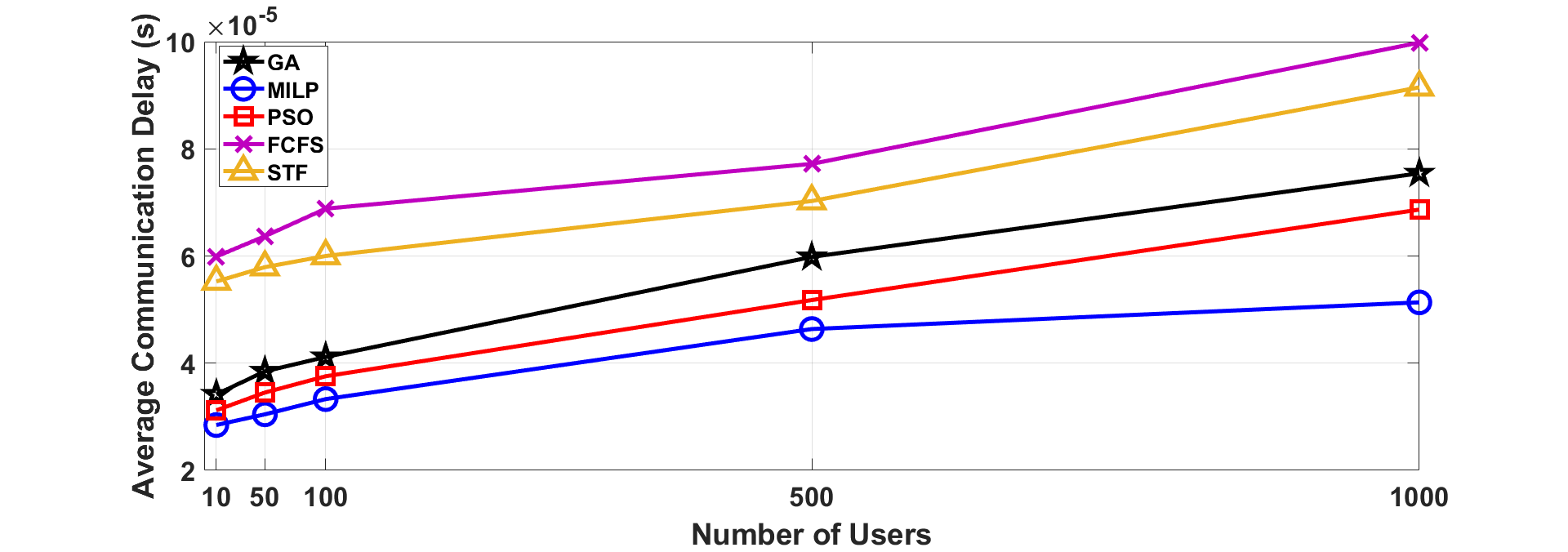}
        \caption{One task per UE}
        \label{fig:comm1 }
    \end{subfigure}
    \begin{subfigure}{0.5\textwidth}
        \centering
        \includegraphics[width =\textwidth, trim= 4.0cm 0cm 3.9cm 0cm ,clip]{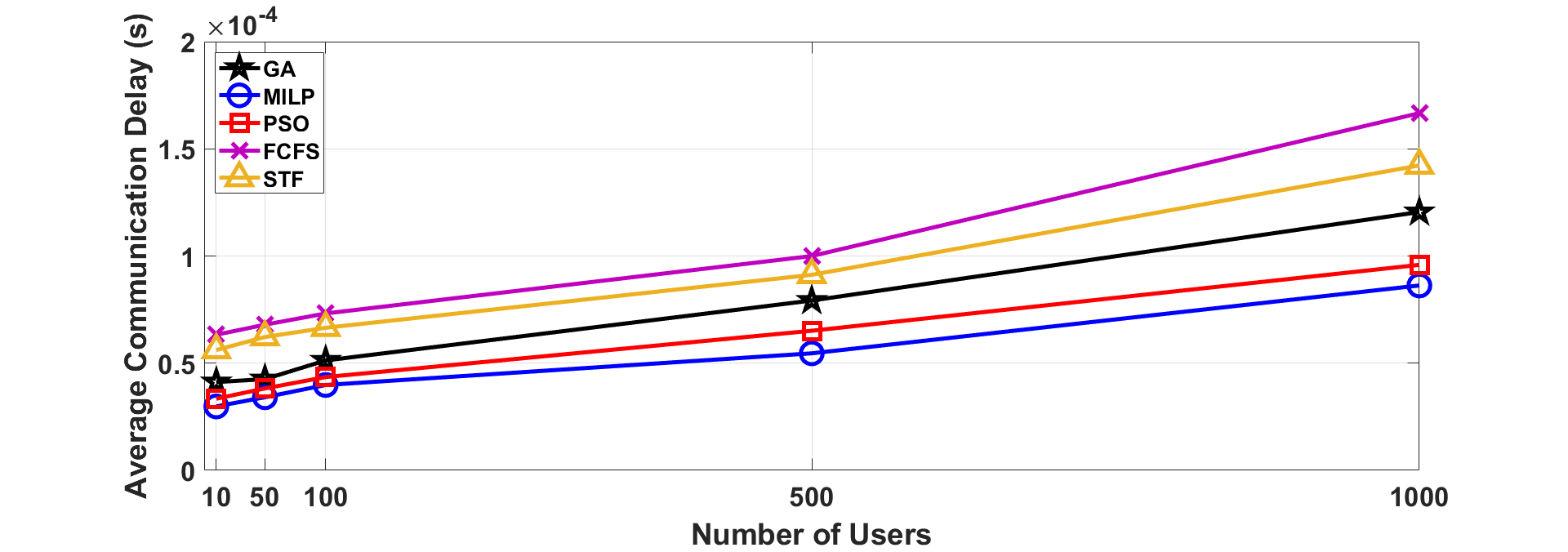}
        \caption{Two task per UE}
        \label{fig:comm2}
    \end{subfigure}
    \begin{subfigure}{0.5\textwidth}
        \centering
        \includegraphics[width =\textwidth, trim= 4.1cm 0cm 3.9cm 0cm ,clip]{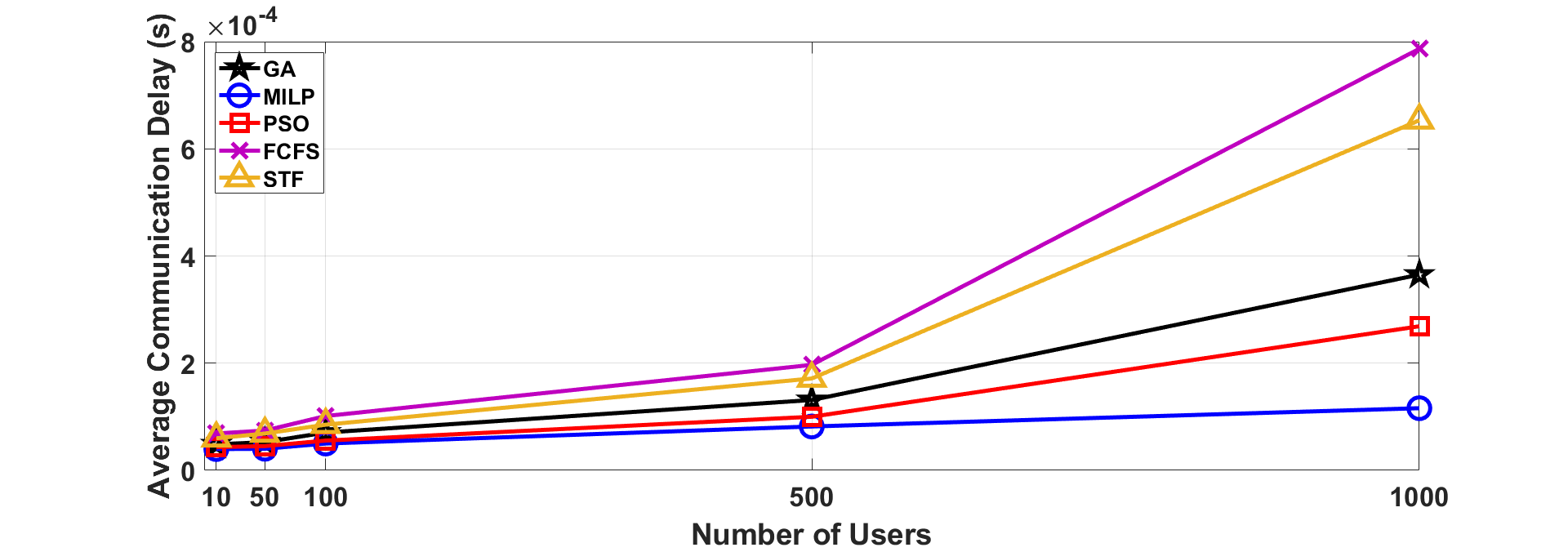}
        \caption{Four task per UE}
        \label{fig:comm4}
    \end{subfigure}
        \caption{Comm. latency for different task counts per UE }
	\label{fig:comm_latency} 
\end{figure}

In all three parts of \figurename \hspace{0.1pt} \ref{fig:convergence} under three different tasks per UE, the MILP   consistently begins with the lowest objective function value. It quickly experiences a sharp decrease and reaches a low and stable value early on in the iterations. This suggests that MILP rapidly identifies a solution that is close to optimal and consistently maintains it, demonstrating robust convergence characteristics.

\begin{figure}[!hbt]
 \centering
    \begin{subfigure}{0.5\textwidth}
        \centering
        \includegraphics[width = \textwidth, trim= 4.2cm 0cm 3.9cm 0cm ,clip]{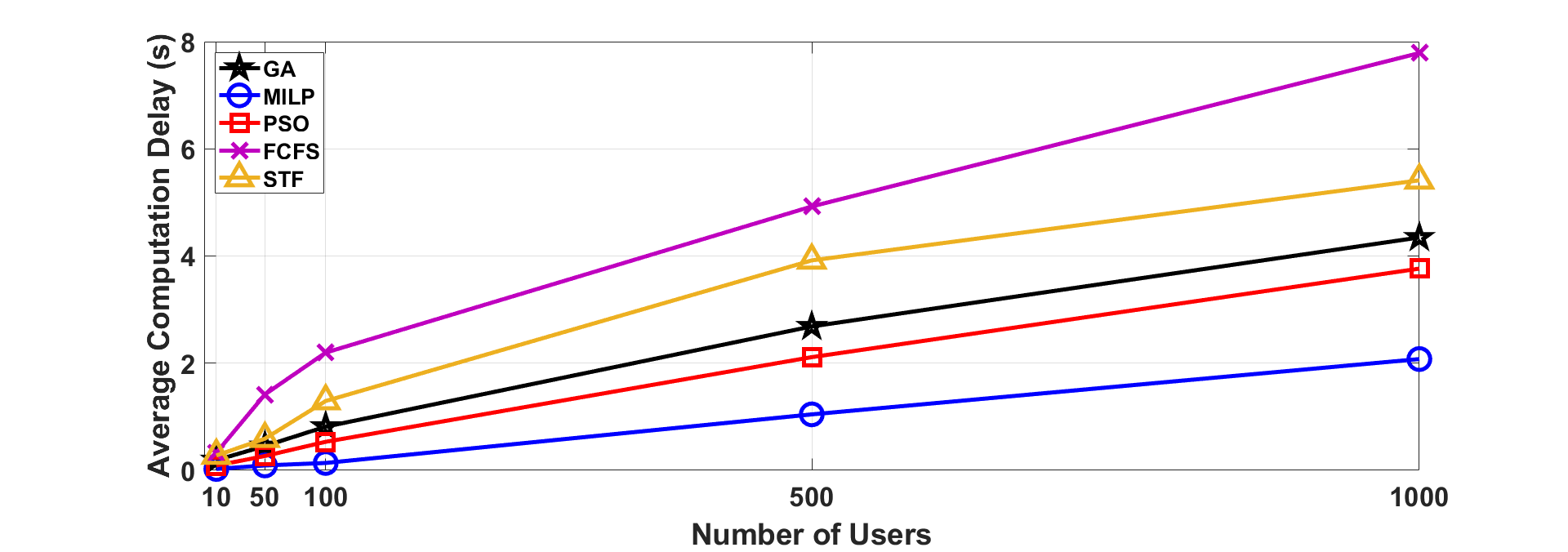}
        \caption{One task per UE}
        \label{fig:comp1}
    \end{subfigure}
    \begin{subfigure}{0.5\textwidth}
        \centering
        \includegraphics[width = \textwidth, trim= 4.2cm 0cm 3.9cm 0cm ,clip]{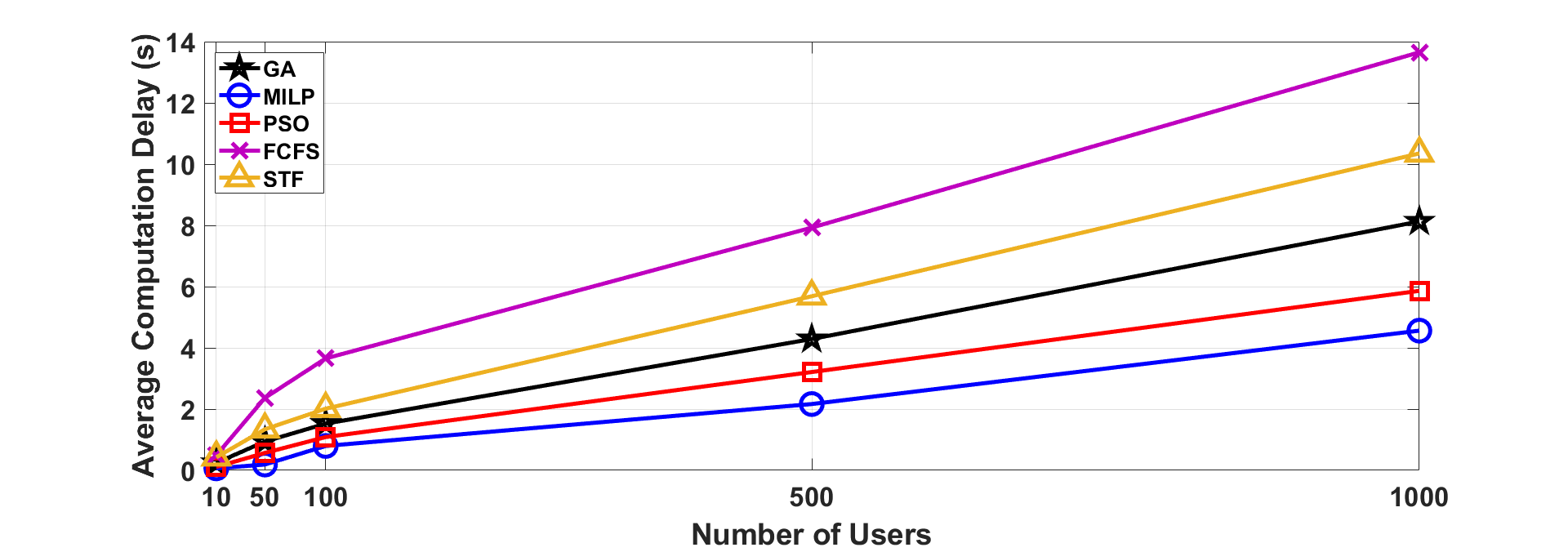}
        \caption{Two task per UE}
        \label{fig:comp2}
    \end{subfigure}
    \begin{subfigure}{0.5\textwidth}
        \centering
        \includegraphics[width = \textwidth, trim= 4.2cm 0cm 3.9cm 0cm ,clip]{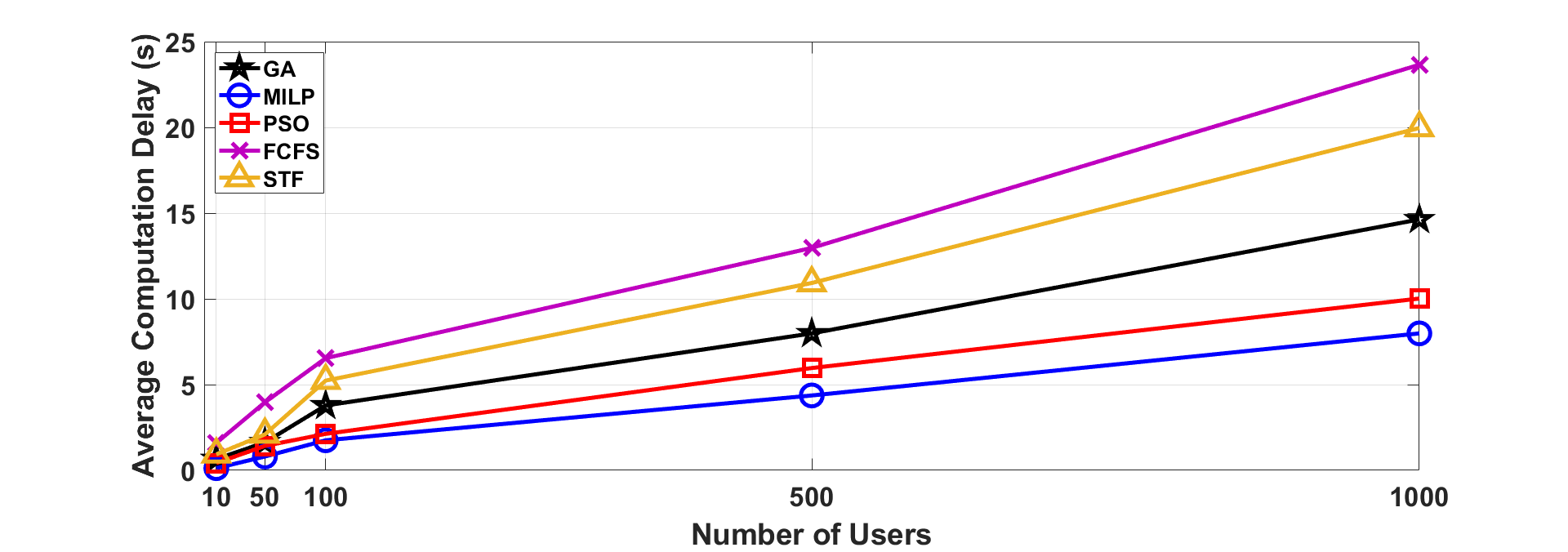}
        \caption{Four task per UE}
        \label{fig:comp4}
    \end{subfigure}
        \caption{Computational latency for different task counts per UE }
	\label{fig:comp_latency} 
 
\end{figure}

\begin{figure}[!hbt]
 \centering
    \begin{subfigure}{0.5\textwidth}
        \centering
        \includegraphics[width =\textwidth, trim= 3.2cm 0cm 3.8cm 0cm ,clip]{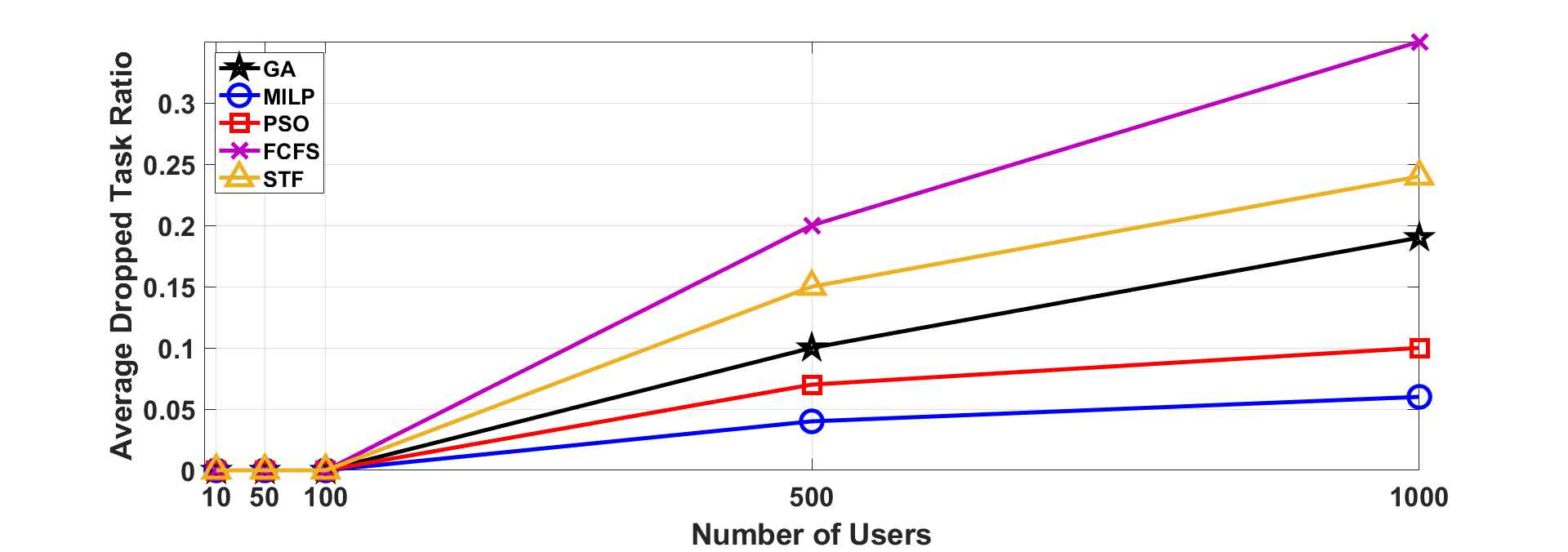}
        \caption{One task per UE}
        \label{fig:dtr1}
    \end{subfigure}
    \begin{subfigure}{0.5\textwidth}
        \centering
        \includegraphics[width = \textwidth, trim= 3.2cm 0cm 3.8cm 0cm ,clip]{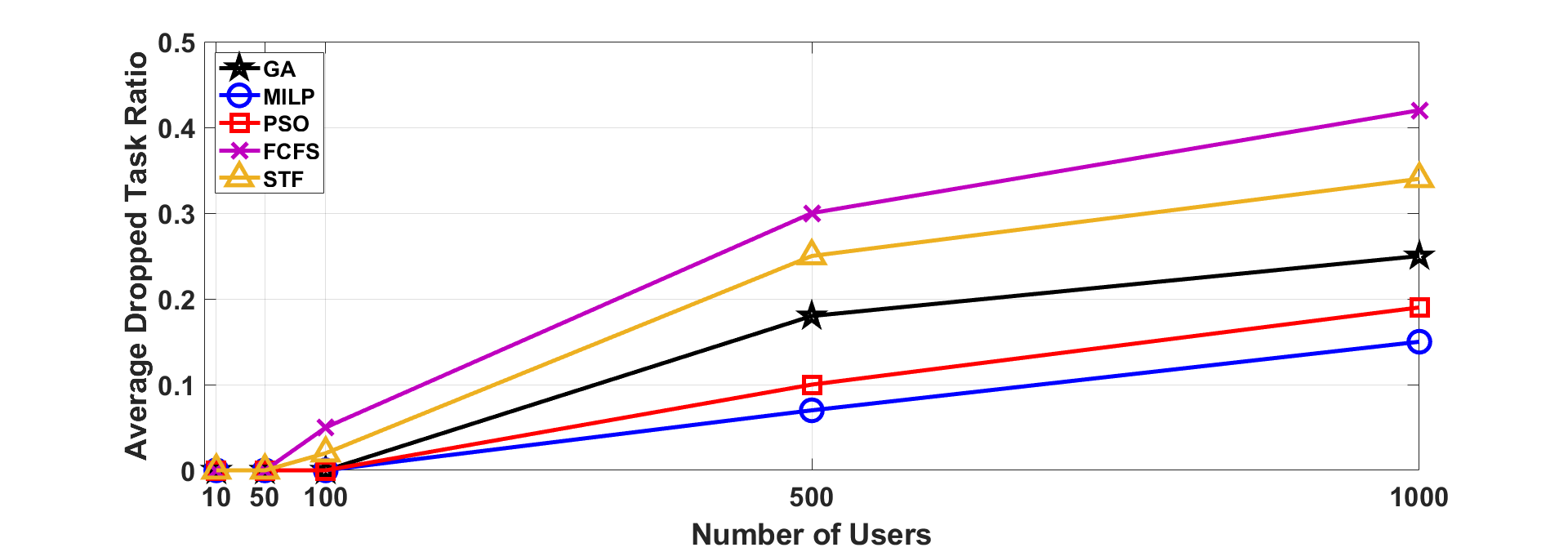}
        \caption{Two task per UE}
        \label{fig:dtr2}
    \end{subfigure}
    \begin{subfigure}{0.5\textwidth}
        \centering
        \includegraphics[width = \textwidth, trim= 3.0cm 0cm 3.8cm 0cm ,clip]{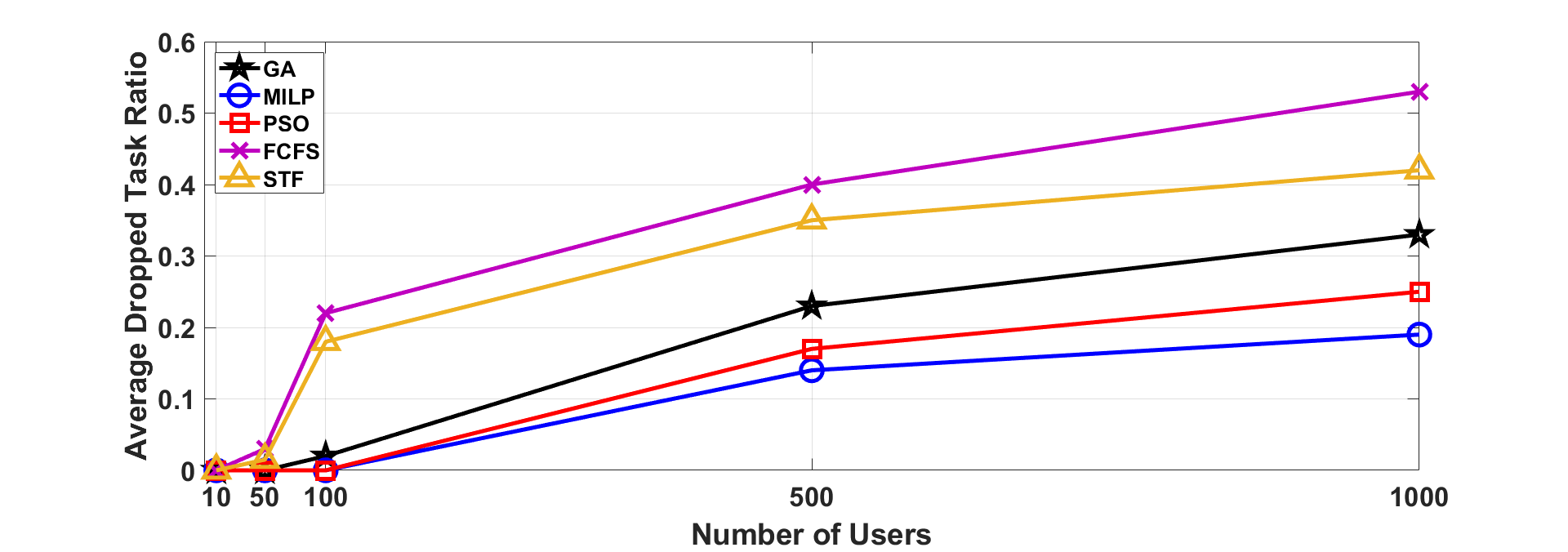}
        \caption{Four task per UE}
        \label{fig:dtr4}
    \end{subfigure}
        \caption{ Dropped task ratio for different task counts per UE }
	\label{fig:dropped_task} 
\end{figure}

\subsubsection{Communication Latency}
Communication latency is related to task size, data rate, and the frequency of the CPU for offloaded tasks.
\figurename \hspace{0.1pt} \ref{fig:comm1 } illustrates a scenario where each UE is assigned only one task. It clearly demonstrates a clear pattern of communication latency growth as the number of UEs increases. FCFS and STF lead to the highest latency when compared to their counterparts. MILP consistently exhibits the lowest average communication latency among different UE numbers. This indicates that MILP is particularly efficient in identifying optimal or nearly optimal solutions to the task offloading problem, since it can linearly represent the problem and systematically explore the feasible region of solution space.

In comparison, the GA demonstrates the greatest mean communication latency across all optimization algorithms. The GA, which emulates biological evolution using probabilistic transition rules, is typically less efficient than MILP. The iterative methodology employed to develop solutions may encounter difficulty in identifying high-quality solutions within a constrained number of generations, particularly when the number of tasks escalates. This can result in less than ideal task allocations, leading to increased latency.

The performance of PSO falls between the performances of MILP and GA. PSO, a heuristic similar to GA, has the advantage of converging to improved solutions faster. This is because the movement of its particles is controlled by both their individual best-known location and the globally best-known position. The dual influence of exploration and exploitation in PSO enables it to effectively navigate the search space, resulting in faster convergence towards optimal solutions compared to GA. Nevertheless, PSO is susceptible to being stuck in local optima, hence hindering its ability to attain the low latency observed under the MILP.

As shown in \figurename \hspace{0.1pt} \ref{fig:comm2}, two tasks per UE, and \figurename \hspace{0.1pt} \ref{fig:comm4}, four tasks per UE, MILP results in the lowest communication latency. The GA demonstrates the highest latency, making it less inefficient compared to MILP and PSO in this particular scenario. In contrast, the MILP approach exhibits minimal latency, indicating a more efficient distribution of tasks. The PSO technique results in moderate latency values, indicating different levels of effectiveness in handling communication overhead. In the multiple-task per UE scenarios, i.e., two and four tasks per UE, the limitations of FCFS and STF become more evident. Under the FCFS scheduling algorithm, an increase in the number of tasks results in longer queues and, thus, greater latency. STF exhibits superior scalability compared to FCFS as the number of tasks increases. The enhanced scalability of STF can be attributed to its prioritization of shorter tasks, resulting in a decrease in the average time a task remains in the system. Although STF successfully reduces latency compared to FCFS by prioritizing tasks, it still faces a substantial latency increase in comparison to GA, PSO, and MILP due to longer tasks' high waiting times behind shorter tasks.

\subsubsection{Computational Latency}
We present the average computational latency for different numbers of tasks per UE in \figurename \hspace{0.1pt} \ref{fig:comp1} which demonstrates that FCFS has its highest latency compared to all other scenarios. When a large task is received immediately prior to a large number of small tasks, the shorter tasks will encounter a latency in their execution as they await the completion of the long task, leading to an increase in their waiting time. 
Conversely, the STF scheduling algorithm prioritizes shorter tasks. When considering task scheduling, it is seen that the arrival of a large task does not hinder the execution of smaller tasks. This leads to a drop in the total waiting time and subsequently reduces computational latency.

The evolutionary technique, albeit robust, is less computationally efficient for this specific task offloading problem, as evidenced by the highest computational latency of GA.
PSO leads to higher latency compared to MILP and lower latency when compared to GA. PSO provides a heuristic-based solution that outperforms GA but falls short of achieving the optimality of MILP. As shown in \figurename \hspace{0.1pt} \ref{fig:convergence}, the PSO algorithm demonstrates a more consistent and steady decline in the value of the objective function when compared to the GA, suggesting a more effective and efficient exploration process. This aligns with the moderate computational latency commonly associated with PSO, presenting it as a compromising solution in terms of optimization speed and computational efficiency. The reason for PSO's decreased latency compared to GA can be attributed to its efficient convergence towards optimal solutions. This is achieved through PSO's ability to quickly converge towards feasible solutions due to its informed search space exploration, guided by both individual and collective particle achievements.

We depict computational latency for two tasks per UE in \figurename \hspace{0.1pt} \ref{fig:comp2} and four tasks per UE in \figurename \hspace{0.1pt} \ref{fig:comp4}. The computational latency increases with the number of tasks per UE. Overall, the convergence graph confirms the fast convergence of MILP in task offloading, since it quickly achieves the lowest values for the objective function, indicating minimum computational latency. PSO, although it operates at a slower pace, outperforms GA by consistently making progress towards achieving lower values. The slower fall of GA is attributed to its significant computing latency, which results in MILP being the most efficient, GA being the least efficient, and PSO performing in between its counterparts.

\begin{figure}[t!]
        \centering
        \includegraphics[width = 0.52\textwidth, trim=0cm 0cm 0cm 0cm,clip]{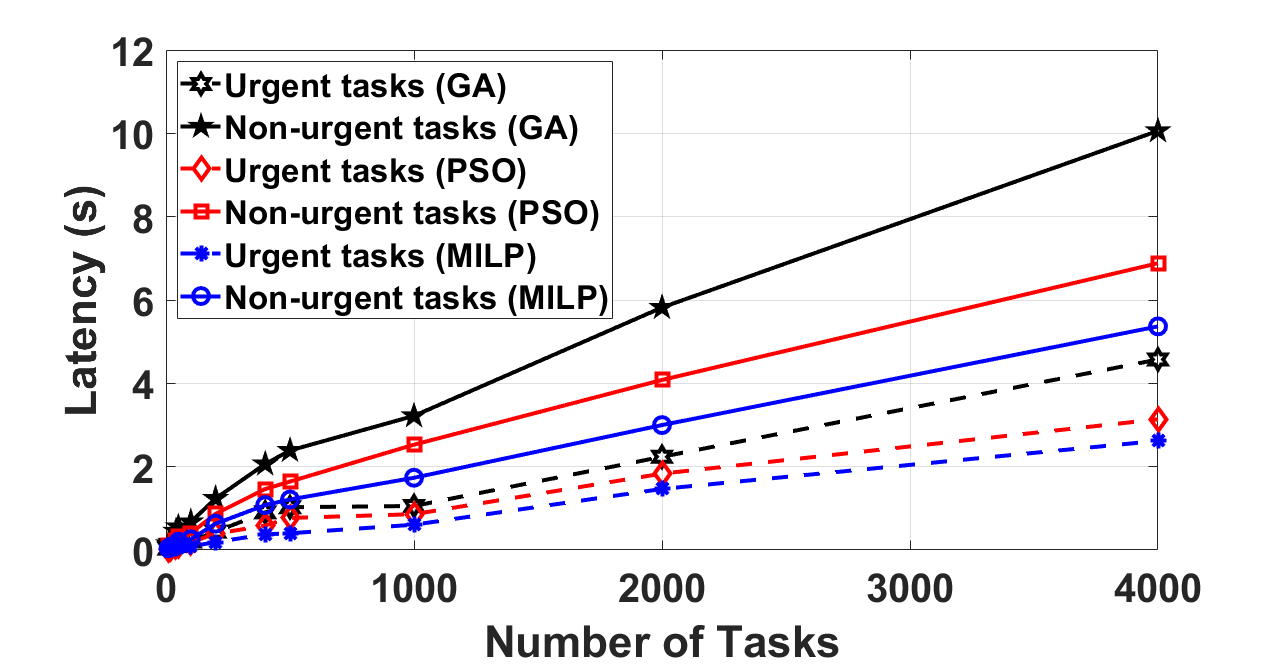}
        \caption{ Urgent UEs vs normal UEs latency }
        \label{fig:urgency} 
\end{figure}

\begin{table*}[!hbt]
\centering
\renewcommand{\arraystretch}{1.5}
\caption{Urgent tasks of UEs VS non-urgent tasks of UEs computational latency for different algorithms}
\label{tab:urgency}
\begin{tabular}{|c|c|c|c|c|c|c|c|c|c|c|}
\hline
Task Type & UE Count & \multicolumn{3}{c|}{One Task per UE} & \multicolumn{3}{c|}{Two Tasks per UE} & \multicolumn{3}{c|}{Four Tasks per UE}\\
\cline{3-11}
 & & GA & PSO & MILP & GA & PSO & MILP & GA & PSO & MILP\\
 \hline
\multirow{5}{*}{Urgent} & 10 & 0.0323 & 0.0199 & 0.0059 & 0.0599 & 0.0345 & 0.0216 & 0.09347 & 0.0767 & 0.0417\\
 & 50 & 0.1709 & 0.0996 & 0.0604 & 0.2051 & 0.1865 & 0.0867 & 0.4401 & 0.3920 & 0.1899\\
 & 100 & 0.2051 & 0.1865 & 0.0867 & 0.4401 & 0.3901 & 0.1893 & 0.9021 & 0.5867 & 0.3742\\
 & 500 & 1.003 & 0.7694 & 0.4021 & 1.0785 & 0.8396 & 0.6158 & 2.2458 & 1.8562 & 1.4456\\
 & 1000 & 1.0561 & 0.8595 & 0.6074 & 2.2373 & 1.8333 & 1.4656 & 4.5724 & 3.1375 & 2.6199\\
 \hline
\multirow{5}{*}{Non-urgent} & 10 & 0.1476 & 0.0998 & 0.0402 & 0.1844 & 0.1179 & 0.0836 & 0.4488 & 0.2322 & 0.1151\\
 & 50 & 0.5533 & 0.3283 & 0.1981 & 0.6695 & 0.3941 & 0.2471 & 1.2213 & 0.8796 & 0.64809\\
& 100 & 0.6725 & 0.4042 & 0.2571 & 1.2371 & 0.8669 & 0.6314 & 2.0631 & 1.4632 & 1.0781\\
& 500 & 2.3887 & 1.6431 & 1.2124 & 3.2196 & 2.5281 & 1.7336 & 5.8195 & 4.0852 & 2.9984\\
& 1000 & 3.1999 & 2.5434 & 1.7158 & 5.8254 & 4.0682 & 4.0856 & 10.0601 & 6.8827 & 5.3663\\
\hline
\end{tabular}
\end{table*}

\begin{figure}[t!]
        \centering
        \includegraphics[width = 0.52\textwidth, trim=0cm 0cm 0cm 0cm,clip]{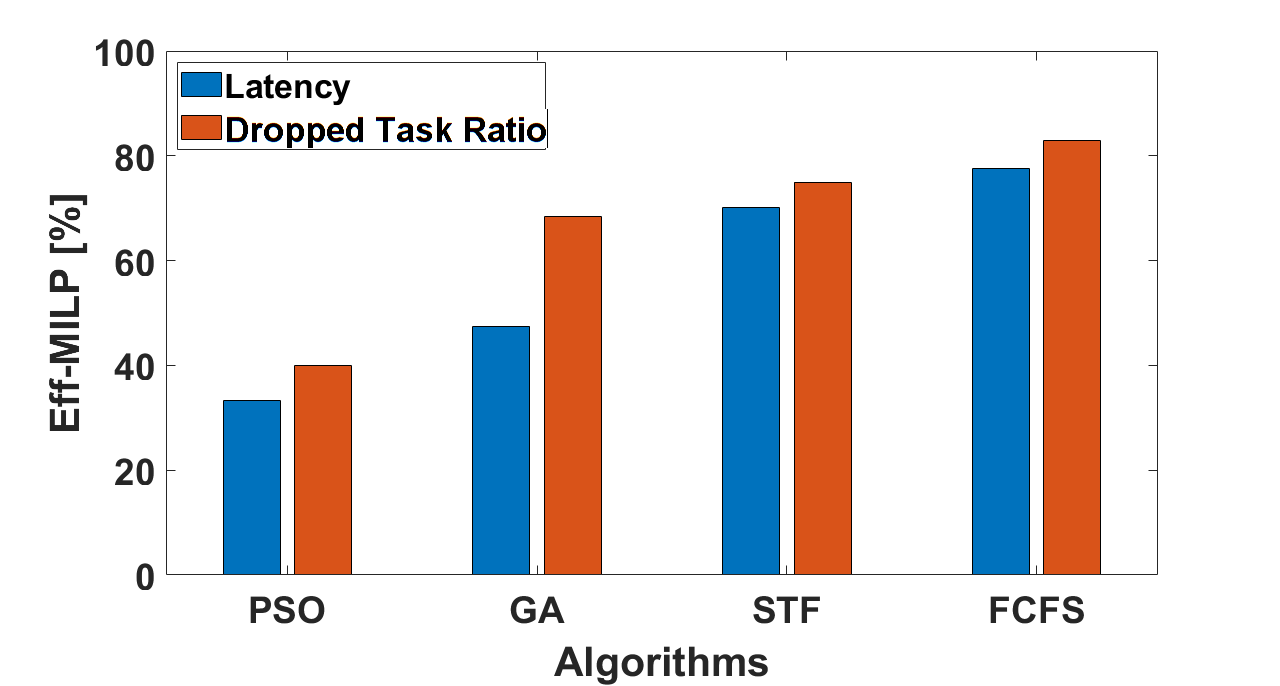}
        \caption{MILP efficiency (\%) in terms of latency and dropped task ratio in comparison to other algorithms for 1000 tasks.}
        \label{fig:MILPefficiency} \vspace{-2mm}
\end{figure}
\begin{figure}[!hbt]
 \centering
    \begin{subfigure}{0.5\textwidth}
        \centering
        \includegraphics[width = \textwidth, trim=4cm 0cm 4cm 0cm,clip]{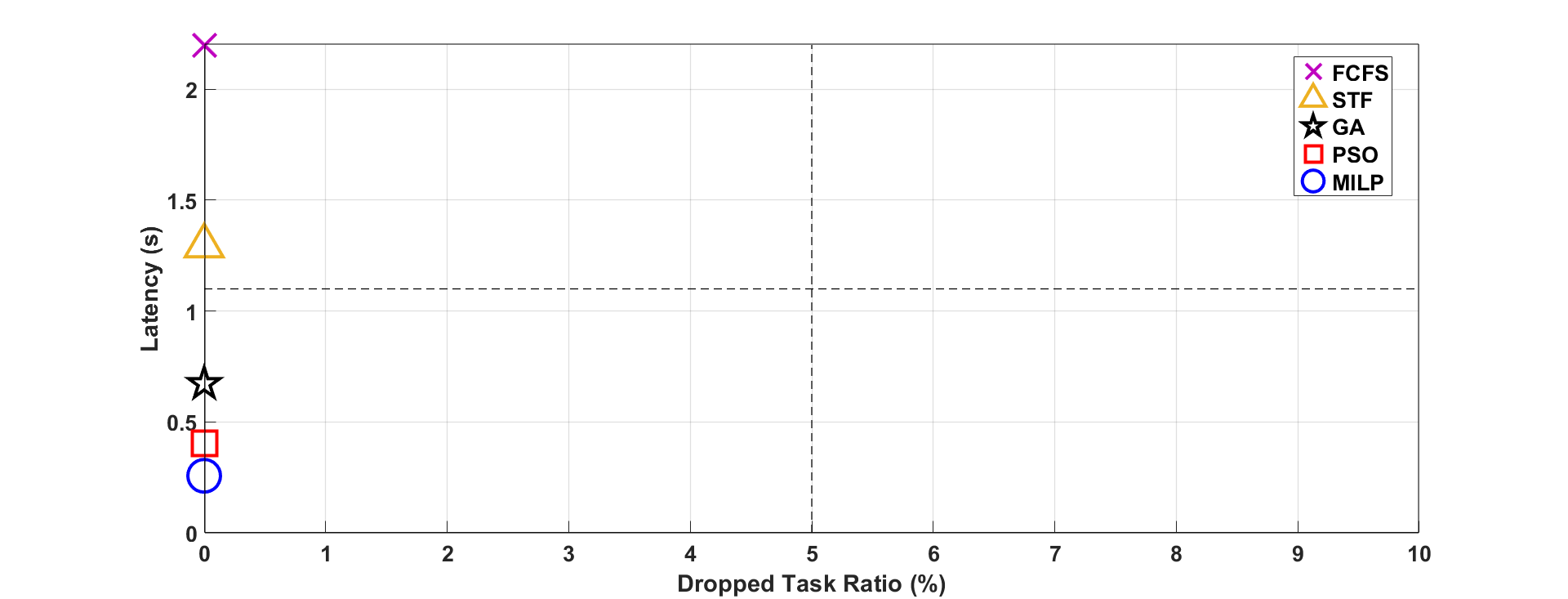}
        \caption{100 non-urgent tasks}
        \label{fig:n100 }
    \end{subfigure}
    \begin{subfigure}{0.5\textwidth}
        \centering
        \includegraphics[width = \textwidth, trim=4cm 0cm 4cm 0cm,clip]{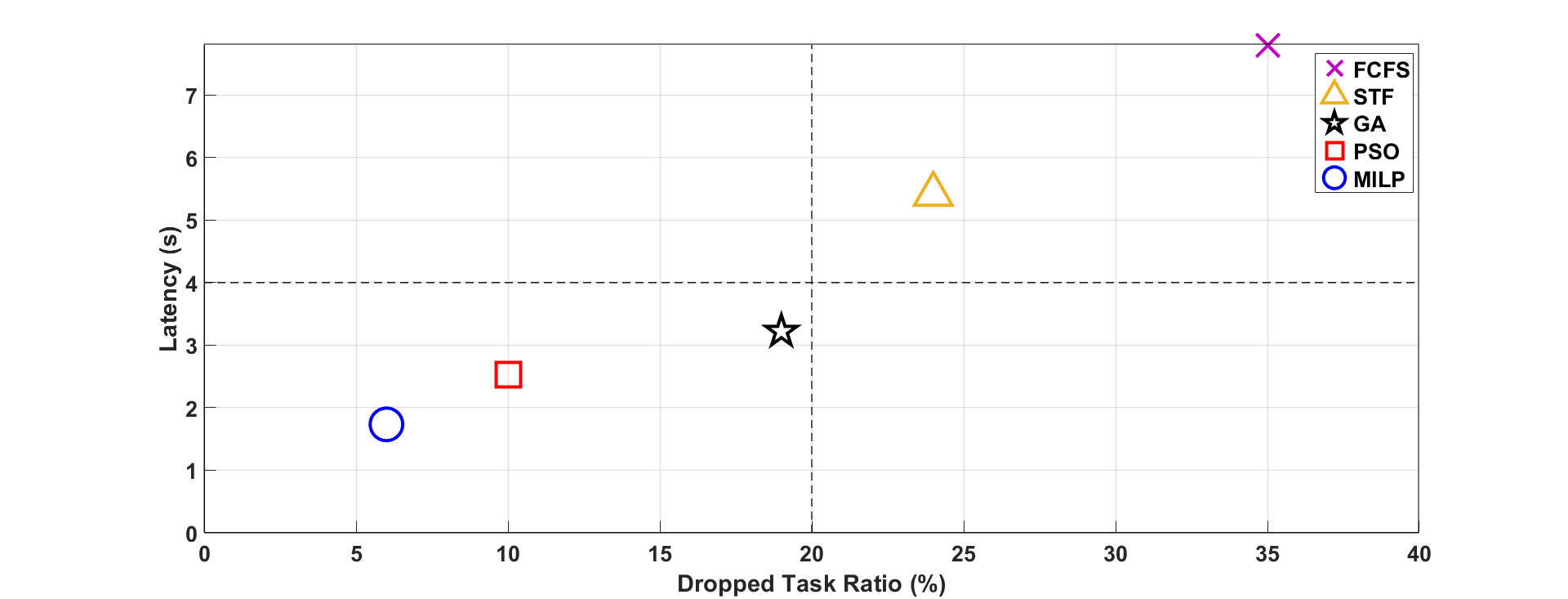}
        \caption{1000 non-urgent tasks}
        \label{fig:n1000}
    \end{subfigure}
    \begin{subfigure}{0.5\textwidth}
        \centering
        \includegraphics[width = \textwidth, trim=4cm 0cm 4cm 0cm,clip]{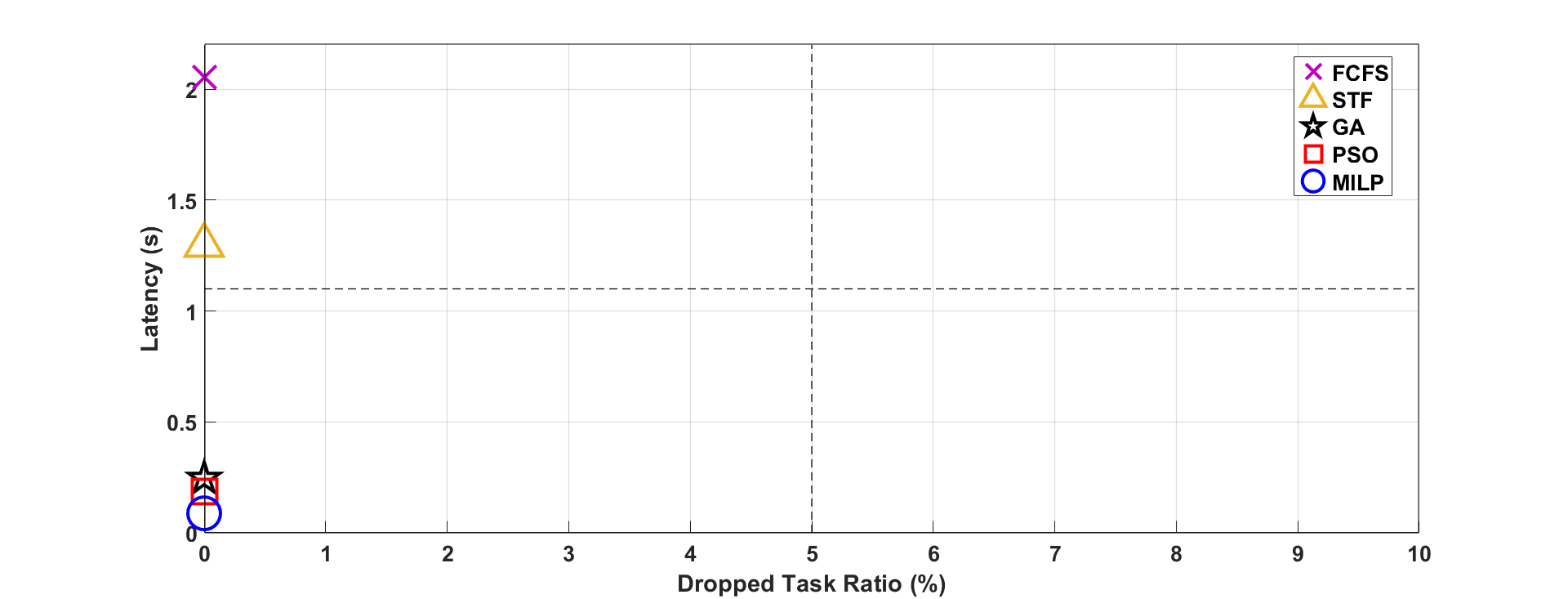}
        \caption{100 urgent tasks}
        \label{fig:u100}
    \end{subfigure}
    \begin{subfigure}{0.5\textwidth}
        \centering
        \includegraphics[width = \textwidth, trim=4cm 0cm 4cm 0cm,clip]{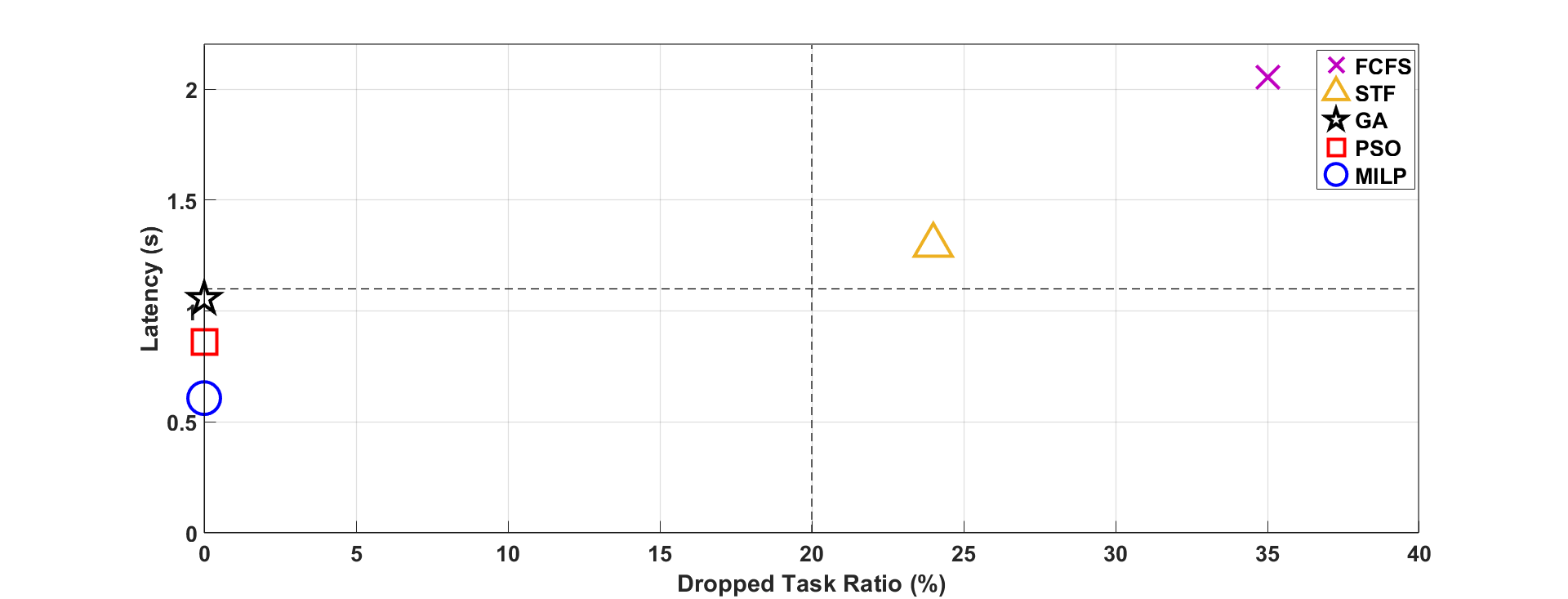}
        \caption{1000 urgent tasks}
        \label{fig:u1000}
    \end{subfigure}
    \caption{Latency and dropped task ratio comparisons}
    \label{fig:ldtr} 
\end{figure}

\begin{table}[h]
\centering
\caption{Total Run Time Comparison}
\newcolumntype{M}[1]{>{\centering\arraybackslash}m{#1}}
\begin{tabular}{|M{1cm}|c|c|c|c|c|}
\hline
\# of Tasks & \multicolumn{5}{c|}{Total Run Time} \\
\cline{2-6} 
 & MILP & PSO & GA & FCFS & STF \\
\hline
10 & 8.10s & 8.39s & 8.71s & 0.12ms & 0.09ms \\
\hline
20 & 8.45s & 8.82s & 9.01s & 0.21ms & 0.17ms \\
\hline
40 & 8.80s & 9.41s & 10.51s & 0.36ms & 0.29ms \\
\hline
50 & 10.95s & 13.65s & 17.39s & 0.48ms & 0.38ms \\
\hline
100 & 3.09min & 4.64min & 5.84min & 0.86ms & 0.71ms \\
\hline
200 & 5.03min & 5.95min & 7.63min & 1.81ms & 1.48ms \\
\hline
400 & 10.57min & 12.93min & 14.04min & 3.96ms & 3.08ms \\
\hline
500 & 14.39min & 15.95min & 18.20min & 5.92ms & 4.79ms \\
\hline
1000 & 21.84min & 25.48min & 28.43min & 13.68ms & 11.97ms \\
\hline
2000 & 38.65min & 47.84min & 52.92min & 29.65ms & 26.84ms \\
\hline
4000 & 85.80min & 95.06min & 114.23min & 65.34ms & 54.58ms \\
\hline
\end{tabular}
\label{tab:run}
\end{table}

\begin{figure}[h]
    \centering
    \includegraphics[width=0.9\columnwidth, trim=0cm 0cm 0cm 0cm, clip]{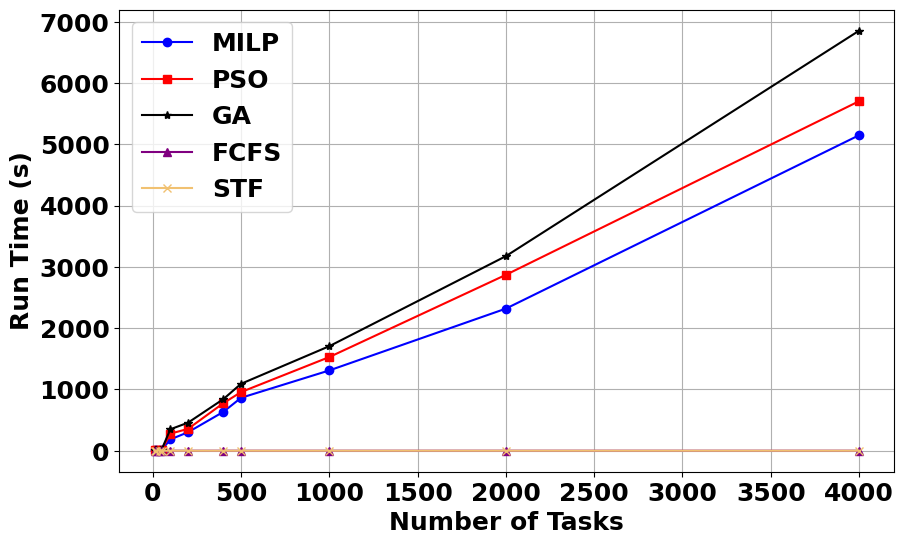}
    \caption{Total Run time for different number of tasks}
    \label{fig:method}
\end{figure}

\subsubsection{Dropped Task Ratio}
As shown in \figurename \hspace{0.1pt} \ref{fig:dtr1} With an increase in the number of UEs, the dropped task ratio tends to increase under all methods, indicating a higher probability of task offloading failures as the system becomes more heavily congested. As shown in \figurename \hspace{0.1pt} \ref{fig:dtr1}, the MILP-based solution exhibits the smallest increase in dropped tasks compared to others. It consistently keeps the dropped task ratio at zero, regardless of the number of UEs. The superior performance indicates that MILP's optimization skills efficiently minimize the number of dropped tasks.

Both GA and PSO in \figurename \hspace{0.1pt} \ref{fig:dtr1} underperform the MILP-based method, with GA showing a slightly higher dropped task ratio compared to PSO. Both algorithms exhibit a more significant increase in the number of dropped tasks compared to MILP as the number of UEs increases. This indicates that while these heuristic-based approaches can adjust to the task offloading problem, they may not consistently identify the most optimal task allocation, resulting in a higher chance of task dropping.

According to \figurename \hspace{0.1pt} \ref{fig:dtr1}, the dropped task ratios for FCFS are higher than STF, showing a significant increase as the number of UEs increases due to the FCFS's processing of tasks in the order of their arrival. In a system with a heavy workload, this might result in inefficiencies, such as lengthy tasks dominating the queue. This can force future tasks to wait for a longer time and raise the chances of them being dropped if they cannot be completed within their specified deadline. By prioritizing smaller tasks, the STF algorithm can generally handle a greater number of tasks within a specific time frame, thereby decreasing the chances of tasks being dropped. STF may accelerate the completion of smaller tasks, allowing for a faster allocation of resources, resulting in a lower dropped task ratio. However, STF is biased in favor of shorter tasks, which can lead to inefficiencies when managing a mixture of small and large tasks during periods of heavy workload, resulting in a high dropped task ratio.
A similar phenomenon is observed in \figurename \hspace{0.1pt} \ref{fig:dtr2} and \figurename \hspace{0.1pt} \ref{fig:dtr4}, where MILP is the most resilient algorithm. On the other hand, heuristic methods such as GA and PSO show intermediary performance, while simpler scheduling approaches like FCFS and STF face challenges in handling heavier demands.

In \figurename \hspace{0.1pt} \ref{fig:MILPefficiency}, efficiency of other algorithms in terms of latency and dropped task ratio compared to MILP is shown. As depicted in \figurename \hspace{0.1pt} \ref{fig:MILPefficiency}, PSO is in second place and more efficient than other algorithms. We calculate the efficiency of each algorithm based on (\ref{eq:e}): 

\begin{equation}
    Eff_{MILP} = \frac{\text{Algorithm Value} - \text{MILP Value}}{\text{Algorithm Value}}
    \label{eq:e}
\end{equation}

\subsubsection{Task Urgency} 

The Table~\ref{tab:urgency} and \figurename \hspace{0.1pt} \ref{fig:urgency} display a comparison of the performance of three optimization algorithms across two types of tasks (urgent and non-urgent) with different numbers of UEs and tasks per UE (one, two, and four tasks per UE). As the number of tasks per UE increases, all algorithms experience an increase in computational latency, regardless of whether the tasks are classified as urgent or non-urgent. The MILP-based solution under all settings achieves better performance than the GA and PSO methods for all UE counts and tasks per UE. It consistently results in the lowest latencies, indicating its ability to obtain the optimal output  results compared to its heuristic or meta-heuristic counterparts in our scenario. Generally, GA has the highest values, suggesting the least efficient performance compared to the other two algorithms. The performance of PSO falls between that of GA and MILP. PSO generally outperforms GA, but it does not reach the same level of efficiency as MILP. This pattern remains consistent across various UE counts and different numbers of tasks.

Table~\ref{tab:urgency} shows a more significant increase for urgent UEs compared to non-urgent ones, particularly as the UE count increases. This implies that the urgency of tasks makes the optimization mathematically harder, hence making the efficient management of tasks more difficult.
As the number of UEs increases from 10 to 1000, the difference in performance between the algorithms becomes more evident, especially between the MILP algorithm and the heuristic algorithms (GA and PSO). This suggests that MILP has superior scalability as the workload increases, enabling more efficient management of tasks as the system becomes more heavily loaded.

In \figurename \hspace{0.1pt} \ref{fig:ldtr}, we depict urgent and non-urgent latency and dropped task ratios for 100 and 1000 tasks. As shown in \figurename \hspace{0.1pt} \ref{fig:ldtr}, FCFS and STF always have a higher latency and dropped task ratio, while MILP, PSO, and GA have the lowest compared to baseline schemes. For 100 tasks (\figurename \hspace{0.1pt} \ref{fig:n100 }) our model achieves a zero dropped task ratio; however, as the number of tasks increases to 1000 (\figurename \hspace{0.1pt} \ref{fig:n1000}, the dropped task ratio increases with latency for non-urgent tasks. The dropped task ratio for urgent tasks (\figurename\ref{fig:u100}  and \figurename \hspace{0.1pt} \ref{fig:u1000} is always zero for MILP, PSO, and GA, with the lowest latency compared to FCFS and STF. In \cite{farooq21}, they analyze computational latency and improve it by an average of 50\% for prioritized users compared to the round robin scheduling algorithm. In our scenario for 100 urgent users, we improve latency in MILP-based by 77\% , PSO by 68\% and GA by 55\% compared to FCFS.

 The table \ref{tab:run} compares total run time (on 4080ti GPU), for each algorithm individually across varying task numbers. For small tasks (10-50), all methods are efficient, with MILP being the fastest. As tasks increase to moderate levels (100-500), MILP maintains lower run times compared to PSO and GA. For large tasks (1000-4000), MILP's efficiency is more pronounced, consistently outperforming PSO and GA. Overall, MILP shows the best performance, making it suitable for larger task sets. Iterative or stochastic algorithms require multiple iterations to explore the solution space and converge toward an optimal solution. However, the deterministic approaches like STF and FCFS follow a predefined rule or sequence to make decisions, resulting in quickly processing tasks without the overhead of exploring multiple solutions. Although they have low computational cost, their results are inefficient compared to optimization algorithms. Additionally, we plot the results in \figurename \ref{fig:method} to show them more clear. 
 
When comparing MILP, PSO, and GA based on computational complexity, MILP worst-case complexity grows exponentially and is typically expressed as $O(2^p \cdot \text{poly}(n))$, where p is the number of integer variables, and poly(n) represents the time to solve the linear relaxation at each branch \cite{rose2023}. The complexity of GA is typically determined by the population size (P), the number of generations (G), and the complexity of evaluating the fitness function for each individual (E), which is usually expressed as $O(G \cdot P \cdot E)$. For PSO, the complexity depends on the number of particles (P), iterations (I), and the complexity of evaluating the fitness function (E), which is often expressed as $O(I \cdot P \cdot E)$ \cite{lach2022}. Despite MILP's exponential complexity, it often runs faster in practice due to its structured techniques like branch-and-bound, which systematically prune the search space. As shown in convergence figure, MILP also converges faster and guarantees an optimal solution for the problem. In contrast, GA and PSO use stochastic search processes, requiring many evaluations and iterations, which lead to longer runtimes as they iteratively explore the solution space. MILP’s deterministic approach ensures exact solutions, whereas GA and PSO typically yield approximate solutions.

\section{Conclusion}
In the context of 5G-MEC, the main challenges related to task offloading are centred around effectively handling the requirement for minimal latency and dropped task ratio, in order to guarantee efficient and scalable system performance. To achieve scalability in the presence of these requirements, it is necessary to implement solutions that optimize task scheduling to effectively manage the large volume of data. We have proposed an approach for task offloading by taking into account computational and communication latency, as well as the dropped task ratio. This approach provides a comprehensive perspective on system efficiency in a realistic and scalable 5G-MEC scenario. The optimization algorithms, MILP, PSO, and GA, have demonstrated superior performance compared to baseline. Under the MILP-based approach, the latency has been decreased by approximately 60\% in comparison to GA and by 55\% in comparison to PSO. The MILP-based approach has reduced the dropped task ratio by approximately 40\% compared to GA and by approximately 20\% compared to PSO. The performance results demonstrate significant efficacy in enhancing service performance within the 5G-MEC framework. 

Our ongoing future work includes incorporating communication and computational energy into these approaches. Furthermore, we are examining multiple scheduling optimization alternatives to enhance the overall efficiency by efficiently utilizing the computational resources at hand and solving challenges in managing increasing demand. Task partitioning to reduce the server workload and latency is also in our agenda. 

\section*{Acknowledgment}
This work was supported in part by funding from the Innovation for Defence Excellence and Security (IDEaS) program from the Department of National Defence (DND), and in part by Natural Sciences and Engineering Research Council of Canada (NSERC) CREATE TRAVERSAL Program.

\normalem
\bibliographystyle{IEEEtran}

\end{document}